\documentclass[twocolumn]{autart}    

\usepackage[utf8]{inputenc}
\usepackage{hyperref}
\usepackage{graphicx}
\usepackage{cite}
\usepackage{wrapfig}
\usepackage{array}
\usepackage{multirow}
\usepackage{enumerate}
\usepackage{amsmath, amscd, verbatim, amsfonts, amssymb, mathrsfs, mathbbol}
\usepackage{graphics, graphicx, xcolor}
\usepackage{kotex}
\usepackage{algorithm}
\usepackage[noend]{algpseudocode}
\usepackage{algorithmicx}
\usepackage{caption}
\usepackage{subcaption}
\usepackage[english]{babel}
\usepackage{bbm}
\usepackage{comment}
\usepackage{upgreek}
\usepackage{tikz}
\usetikzlibrary{arrows}
\usetikzlibrary{arrows.meta,positioning}
\usepackage{arydshln}
\usepackage{scalerel}
\usepackage[labelformat=simple]{subcaption}

\DeclareMathOperator*{\argmax}{argmax}
\DeclareMathOperator*{\argmin}{argmin}

\DeclareMathOperator{\diag}{\text{diag}}
\DeclareMathOperator{\minimize}{\text{minimize}}
\DeclareMathOperator{\st}{\text{subject to}}

\newcommand{\ifnote}[1]{} 

\newcommand{\VEC}[1]{\text{vech} \left({#1}\right)}
\newcommand{\invVEC}[1]{\text{vech}^{-1}({#1})}
\newcommand{\msf}[1]{\mathsf {#1}}

\newcommand{\blue}[1]{{\color{blue}#1}}
\newcommand{\orange}[1]{{\color{orange}#1}}

\newcommand{\augH}{{\bf H}}
\newcommand{\augS}{{\bf S}}
\newcommand{\augz}{{\bf z}}
\newcommand{\diagg}[1]{\bar{#1}}

\newcommand{\augE}{\mathscr E}
\newcommand{\augU}{\mathscr U}
\newcommand{\augF}{\mathscr F}
\newcommand{\augG}{\mathscr G}
\newcommand{\augHall}{\mathscr H}
\newcommand{\augRall}{\mathscr R}
\newcommand{\augP}{\mathscr P}

\hypersetup{colorlinks,plainpages,linkcolor=black,citecolor=black, urlcolor=black}

\begin{document}

\begin{frontmatter}

\title{Consensus optimization approach for distributed {{K}}alman filtering: performance recovery of centralized filtering with proofs 
} 

\thanks[footnoteinfo]{A preliminary version of this article was presented at the 58th IEEE Conference on Decision and Control \cite{Ryu2019CDC}.
}

\author[KWU]{Kunhee Ryu}\ead{ryuhhh@kw.ac.kr},    
\author[KWU]{Juhoon Back}\ead{backhoon@kw.ac.kr}               

\address[KWU]{School of Robotics, Kwangwoon University, Seoul 01890, Korea
}  


\begin{abstract}                          
This paper investigates the distributed Kalman filtering (DKF) from distributed optimization viewpoint.
Motivated by the fact that Kalman filtering is a maximum a posteriori estimation (MAP) problem, which is a quadratic optimization problem, we reformulate DKF problem as a consensus optimization problem, resulting in that it can be solved by many existing distributed optimization algorithms. A new DKF algorithm employing the dual ascent method is proposed, and its stability is proved under mild assumptions. 
The performance of the proposed algorithm is evaluated through numerical experiments. \vspace{-0.4cm}
\end{abstract}
\end{frontmatter}
\section{Introduction}\vspace{-0.2cm}
In order to monitor large scale systems or environments such as traffic networks, plants, sea, etc., distributed filtering using multiple estimators is preferred because it has advantages in terms of scalability, robustness to component loss, and computational cost. Although promising, developing fully distributed solutions with guaranteed stability and optimality is still challenging due to practical issues including heterogeneity of sensors, restriction on communication, uncertainty of network topology, etc. Against this backdrop, literature on distributed Kalman filtering (DKF) is expanding rapidly \cite{Berg1994ACC, Olfati2005CDC, Bai2011ACC, Kim2016CDC, Liu+2022TAC, Deshmukh2017optimal, Li2015distributed, Yan+2022TAC, Carli2008distributed, Yu+2019TSMC:S}; see also the survey \cite{Mahmoud2013Survey} and references therein.

Recently, consensus based DKF algorithms have received particular attention since the seminal works \cite{Olfati2005CDC, Olfati2007CDC, Olfati2009CDC} have been published. In \cite{Olfati2005CDC}, the author addressed the connection between the consensus problems and DKF problems. Two DKF algorithms, called Kalman-Consensus filters have been presented in \cite{Olfati2007CDC}. In the first algorithm, each filter calculates the average of the measurements across all filters in a distributed way and then updates local estimate using it, while in the other, local estimates are obtained by applying  standard Kalman filtering computation and then the filters draw consensus on these estimates. See the works \cite{Olfati2007CDC, Olfati2009CDC,KamgarpourTomlin2008CDC} for the analysis on stability and optimality.

Notably, in \cite{Battistelli2015TAC}, three average consensus based algorithms (called consensus on information, consensus on measurement, and their hybrid type) have been presented for collectively observable sensor networks. These algorithms perform subiterations to compensate for insufficient information from the unobservable subspace of individual sensor and to accelerate consensus. 
The consensus on measurement algorithm, for example, finds the weighted averages on measurements and information rate matrices (see Section \ref{sec:problemsetting} for the definition) through the consensus step (subiteration), and by using them, the estimates and covariances are corrected.
Motivated by these works, various DKF algorithms have been developed; algorithms with consistency \cite{BattistelliChisci2014Aut, He+2018Aut} and asymptotic optimality \cite{Battilotti+2021Aut, Battilotti+2020Aut}, and those for collectively detectable sensor networks \cite{Nozal+2019Aut, Li+2020TAC, Wei+2020SMC:S}. 
There are also alternative approaches which are not directly connected to average consensus; e.g., diffusive DKF \cite{CattivelliSayed2010TAC} and the dynamic consensus on pseudo-observation DKF \cite{DasMoura2015TSP}.
See also the works on distributed Kalman-Bucy filtering  \cite{Kim2016CDC, Ren2017SJ}, partition-based DKF \cite{Farina2016TCNS}, DKF with state constraints \cite{He2019TAC}, DKF for uncertain systems \cite{Zorzi2019TCNS}, etc.

To the best of the authors' knowledge, most of the researches on DKF try to find a good way to fuse the outcomes of local Kalman filters, and this is done by modifying the filter structure or adding extra consensus procedures. Although many noteworthy results have been produced so far, a more fundamental problem formulation of DKF, which naturally embraces the structural constraint from the communication network and admits an optimal solution, is still missing.

In this paper, we reconsider the DKF problem from the distributed optimization perspective, motivated by the fact that Kalman filtering is basically an optimization problem \cite{Mendel1995PH, Chen2003Statistics, Bell1993TAC, Thrun2005Book}. Under the assumption that the measurement noises of sensors are mutually uncorrelated, we observed that the cost function of centralized Kalman filtering (CKF) can be decomposed into parts so that each part depends on only one sensor. From this, it is shown that the DKF problem can be reformulated as a consensus optimization problem \cite{Boyd+2011FTML}, which is the first contribution.

One important implication of the reformulation is that novel DKF algorithms employing distributed optimization methods can be developed. In view of this, a new DKF algorithm is proposed in this paper by employing the dual ascent method, which is the second contribution. It is noted that the proposed algorithm is an improved version of \cite{Ryu2019CDC} in the sense that less information is required to choose estimator gains. 
In addition, it is proved that the proposed algorithm is unbiased, and that the covariance of each estimator converges to the steady-state covariance of CKF \cite{Battilotti+2020Aut, Battilotti+2021Aut}. This ensures that the proposed algorithm asymptotically recovers the performance of CKF. The stability analysis is done under standard assumptions on the target system commonly made in the Kalman filtering \cite{Jazwinski2007Book, Kamen1999Book}. 
Precisely, we assume neither the local observability of the sensor network \cite{DasMoura2015TSP, CattivelliSayed2010TAC, Olfati2009CDC} nor the invertibility of the system matrix \cite{Battistelli2015TAC, Battistelli2016Aut, He+2018Aut, He2019TAC, Duan2020Aut}. This contributes to the theoretical completeness of this study.

This paper is organized as follows. In  Section \ref{sec:problemsetting}, we connect DKF problem to a distributed optimization problem. A new DKF algorithm based on the dual ascent method is proposed in Section \ref{sec:DA_DKF}, and stability analysis is presented in Section \ref{sec:stability}. Numerical experiments are given in Section \ref{sec:numerical_experiments}.
Section \ref{sec:conclusion} concludes the paper.

\noindent{\bf Notation}: 
For matrices $A_1, \dots, A_n$, $\diag\{A_1, \dots, A_n\}$ denotes the block diagonal matrix composed of $A_1, \dots, A_n$. For vectors $a_1,\dots,a_n$, $[a_1;\cdots;a_n] := [a_1^\top,\cdots,a_n^\top]^\top$, and $[A_1;\cdots;A_n]$ with matrices $A_i$'s defined similarly. $1_n \in \mathbb{R}^n$ denotes the vector whose components are all 1, and $I_n \in \mathbb R^{n\times n}$ and $0_{n \times n} \in \mathbb R^{n\times n}$ are the identity matrix and the zero matrix, respectively. 
For a symmetric matrix $A$, $A>0$ ($A\ge 0$, resp.) denotes that $A$ is a positive definite (semidefinite, resp.) matrix. 
We write $x \sim \mathsf{N}(\mu, \sigma^2)$ when $x$ is normally distributed with mean $\mu$ and variance $\sigma^2$. $\mathbb{E}\{x\}$ and  ${\rm Cov}(x)$ denote the expectation and covariance of  $x$, respectively.
For a symmetric matrix $M \in \mathbb{R}^{n \times n}$, $\VEC{M} \in \mathbb{R}^{n(n+1)/2}$ denotes the half vectorization of $M$, a column vector obtained by using only the upper triangular part of $M$, and $\invVEC{\cdot}$ denotes the inverse of $\VEC{\cdot}$. 
For a function $f(x, y)$, $\nabla_{x} f(x,y)$ denotes the gradient vector with respect to $x$.
$\delta_{kl}$ represents the Kronecker delta. 

\vspace{-0.3cm}
\section{Distributed Kalman filtering and its Connection to Consensus Optimization}\vspace{-0.2cm}\label{sec:problemsetting}
Consider a linear system with $N$ sensors given by
\begin{subequations}\label{eq:System}
\begin{align}
x_{k+1} &= Fx_k + w_k \label{eq:System_state}\\
y_k &= H x_k + v_k =  \begin{bmatrix} H_1 \\ \vdots \\ H_N \end{bmatrix} x_k + \begin{bmatrix} v_{1,k} \\ \vdots \\ v_{N,k} \end{bmatrix} \label{eq:System_output}
\end{align}
\end{subequations}
where $x_{k} \in {\mathbb{R}}^{n}$ is the state vector, $y_k := [y_{1,k};\cdots;y_{N,k}] \in \mathbb{R}^{m}$ is the measurement vector, and $y_{i,k} \in \mathbb{R}^{m_i}$ is the measurement associated with sensor $i$ where $m_i$'s satisfy $\sum^N_{i=1} m_i = m$. $F$ is the system matrix and $H$ is the output matrix consisting of $H_i \in \mathbb{R}^{m_i \times n}$ that is the output matrix associated with sensor $i$. 
The process noise is denoted by $w_k$ and $v_{i,k}$ is the measurement noise on sensor $i$, which are zero-mean Gaussian. $w_k$ and $v_{i,k}$ are mutually uncorrelated jointly Gaussian and white, i.e., $\mathbb E\{ w_k w_l^\top\} = Q\delta_{kl}$, $\mathbb E\{ v_{i,k} v_{j,l}^\top\} = R_i\delta_{ij}\delta_{kl}$, and  $\mathbb E\{ w_k v_{i,l}^\top \} = 0$ for any $i,j = 1,\dots,N$ and positive integers $k,l$. It is assumed that $Q>0$ and $\bar R := \diag\{R_1, \dots, R_N\} > 0$. In addition, let the initial state vector $x_0$ be Gaussian, with mean $\mathbb E\{x_0\}$ and covariance $P_0 > 0$, i.e., $x_0 \sim \mathsf N(\mathbb E\{x_0\}, P_0)$. It is supposed that $x_0$ is uncorrelated with $w_k$ and $v_k$.

\begin{assum}\label{as:collecObs}
    The pair $(F, H)$ is observable.
\end{assum}

\begin{rem}
Assumption \ref{as:collecObs} means that it may not be possible to estimate the state of the system using the measurement from a single sensor, i.e., $(F,H_i)$ may not be observable, while the whole sensor network consisting of $N$ sensors satisfies the usual sense of observability.  
\end{rem}

If all the measurements from $N$ sensors are collected and processed altogether, the problem can be seen as the one with an imaginary sensor that measures $y_k$ with knowledge on $H$, thus called centralized Kalman filtering. 
The filtering consists of two steps, {\it{prediction}} and {\it{correction}}, and it is well known that the update rules can be derived from MAP (maximum a posteriori) approach in the Bayesian framework. 

Let $Y_k:= [y_0;\cdots;y_k]$. In the view of MAP \cite{Mendel1995PH}, the optimal estimate of Kalman filtering is defined as $\hat x_{k} = \argmax_{x_k} p(x_k|Y_k)$. 
Suppose that $\hat x_{k-1} = \mathbb E\{x_{k-1}|Y_{k-1}\}$ and $P_{k-1} = {\rm Cov}(x_{k-1}|Y_{k-1})$ be the estimate of the state and its covariance at preceding time are given. In the prediction step, the predictive estimate $\hat x_{k|k-1}$ and covariance matrix $P_{k|k-1}$ are computed as $\hat x_{k|k-1} = F \hat x_{k-1}$ and $P_{k|k-1} = F P F_{k-1} F^\top + Q$.

By defining $\augz_{c,k} = [y_k;\hat{x}_{k|k-1}]$, $\augH_c = [H;I_n]$, $\augS_{c,k} =\text{diag}\{\diagg{R}, P_{k|k-1}\}$, the optimal estimate can be equivalently obtained as $\hat x_k = \argmin_{\xi_{c,k}} f_{c,k}(\xi_{c,k})$ where $\xi_{c,k} \in \mathbb R^n$ is the free variable and the cost function is given by
\begin{align}\label{eq:f_ck}
    f_{c,k}(\xi_{c,k}) = \frac{1}{2}(\augz_{c,k}-\augH_c \xi_{c,k})^{\top} \augS_{c,k}^{-1} (\augz_{c,k}-\augH_c \xi_{c,k}).
\end{align}
Since $f_{c,k}(\xi_{c,k})$ is a convex function, provided that $P_{k|k-1} > 0$, $\hat x_k$ can be obtained from
$\nabla_{\xi} f_{c,k}(\hat x_k) = 0$, from which we have the correction step as 
\begin{align*}
\hat x_k &= \hat x_{k|k-1} + K_k (y_k - H\hat x_{k|k-1})\\
P_k &= P_{k|k-1} - P_{k|k-1} H^\top (H P_{k|k-1} H^\top + \diagg{R})^{-1} H P_{k|k-1}.  
\end{align*}
where $K_k = (H^\top \diagg{R}^{-1} H + P^{-1}_{k|k-1})^{-1} H^\top \diagg{R}^{-1}$ that is the Kalman gain. For further details, see, e.g., \cite{Mendel1995PH, Chen2003Statistics, Kamen1999Book, Jazwinski2007Book}.

Now we consider the DKF problem.
Each estimator in the network tries to find the optimal estimate by processing the local measurement and exchanging information with its neighbors. The communication network among estimators is modeled by a graph $\mathcal{G}$, and $\mathcal N$ and $\mathcal N_i$ denote the node set and the neighbor set of estimator $i$, respectively. The Laplacian matrix associated with $\mathcal{G}$ is denoted by $L \in \mathbb{R}^{N \times N}$ and $a_{ij}$ is a weight of the edge between nodes $i$ and $j$. It is known that $L$ has a simple zero eigenvalue corresponding to the unit eigenvector $\mathsf u = \frac{1}{\sqrt{N}} 1_N$, and there exists an orthogonal matrix $U := [\mathsf u ~ W]$ such that $LU = U\Lambda$ where $W \in \mathbb R^{N \times (N-1)}$ is a matrix consisting of unit eigenvectors corresponding to the nonzero eigenvalues of $L$, denoted by $\sigma_2, $\dots$, \sigma_N$, and $\Lambda = \diag\{0,\tilde \Lambda\}$ with $\tilde \Lambda = \diag\{\sigma_2, \dots, \sigma_N\}$. To proceed, we define the following to simplify the notation.
\begin{align*}
{\mathbb 1}_N&=1_N \otimes I_n,\ {\mathbb I}_N = I_N \otimes I_n, \ {\mathbb U}=U\otimes I_n, \ 
{\mathbb W} = W\otimes I_n\\
{\mathbb L} &= L\otimes I_n, \ \ \quad\!\! {\mathbb \Lambda}= \Lambda \otimes I_n, \   \quad\!\! \tilde {\mathbb \Lambda}= \tilde \Lambda \otimes I_n. 
\end{align*}
For the network, we make the following assumption.
\begin{assum}\label{as:network}
The network $\mathcal{G}$ is undirected and connected, and the maximum eigenvalue of $L$, denoted by $\sigma_N$, is bounded by $\bar \sigma$ which is known. 
\end{assum} 

Under the setting (\ref{eq:System}), estimator $i$ acquires only the local measurement $y_{i,k}$, and the parameters $H_i$ and $R_i$ are kept private to estimator $i$. It is noted that the pair $(F, H_i)$ is not necessarily observable, and we assume that $F$ and $Q$ are open to all estimators, and $N$ is known.

Similar to CKF, DKF is performed in two steps, local prediction and distributed correction. In the local prediction step, each estimator predicts
\begin{equation*}
\hat{x}_{i, k|k-1} = F \hat{x}_{i, k-1}, ~ 
P_{i, k|k-1} = F P_{i, k-1} F^\top + Q
\end{equation*}
where $\hat{x}_{i, k|k-1}$ and $P_{i, k|k-1}$ are local estimates of $\hat{x}_{k|k-1}$ and $P_{k|k-1}$, respectively, that estimator $i$ holds.

The distributed correction step solves the MAP estimation problem in a distributed manner.  First, we define
$\augz_{i,k} = [y_{i,k};\hat{x}_{i,k|k-1}]$, $\augH_i = [H_i; I_n]$, and $\augS_{i,k} = \diag\{R_{i}, N P_{i,k|k-1}\}$ which depend on only local variables and parameters.
Owing to the structure of ${\bf S}_{c,k}$, the cost function $f_{c,k} (\xi_{c,k})$ in \eqref{eq:f_ck} can be decomposed into 
    $f_{c,k}(\xi_{c,k}) = \sum_{i=1}^N f_{i,k} (\xi_{c,k})$
where $f_{i,k} (\xi_{c,k}) = \frac{1}{2} (\augz_{i,k} - \augH_i \xi_{c,k})^\top \augS_{i,k}^{-1} (\augz_{i,k} - \augH_i \xi_{c,k})$. It is noted that the cost becomes identical to that of CKF when the estimators reach a consensus on $\hat{x}_{i, k-1}$ and $P_{i, k-1}$ in the correction step at time $k-1$.

Allowing that each estimator holds its own optimization variable $\xi_{i,k} \in \mathbb{R}^{n}$ for $\xi_{c,k}$, DKF problem becomes a consensus optimization problem given by
\vspace{-0.2cm}
\begin{equation*}\tag{P.1}\label{p:estimate}
\begin{split}
    \underset{\xi_{1,k}, \dots, \xi_{N,k}}{\text{minimize}}& \quad \sum_{i=1}^N f_{i,k}(\xi_{i,k})\\
    \text{subject to}& \quad \xi_{1,k} = \cdots = \xi_{N,k}.
\end{split}
\end{equation*}
If there exists a distributed algorithm that finds a minimizer, we say that the algorithm solves DKF problem. 

Since the kernel of Laplacian $L$ is $\text{span} \{1_N\}$, the constraints of \eqref{p:estimate} can be written as $\mathbb L \xi_k = 0$ where $\xi_k = [\xi_{1,k};\cdots;\xi_{N,k}]$. We define the Lagrangian for \eqref{p:estimate} as 
\begin{align}\label{eq:LagrangianDKFP}
\mathcal{L}_{\mathsf{est},k} (\xi_k, \lambda_k) &= \sum_{i=1}^N f_{i,k}(\xi_{i,k}) + \lambda^\top_k \mathbb L \xi_k
\end{align}
where $\lambda_k \in \mathbb{R}^{Nn}$ is the Lagrange multiplier (dual variable) associated with the consensus constraint. 
Define 
\begin{equation}\label{eq:define_z_H_S}
\begin{split}
\augz_k &= [\augz_{1,k};\cdots;\augz_{N,k}], \diagg{\augH} =\diag\{\augH_1, \dots, \augH_N\} \\
\diagg{\augS}_k &= \diag\{\augS_{1,k}, \dots, \augS_{N,k}\}, \mathcal{H}_k = {\mathbb 1}_N^\top \diagg{\augH}^\top \diagg{\augS}_k^{-1} \diagg{\augH} {\mathbb 1}_N.
\end{split}
\end{equation}
Note that $\mathcal H_k$ is a symmetric positive definite matrix.

We rewrite the Lagrangian (\ref{eq:LagrangianDKFP}) as $\mathcal{L}_{\mathsf{est},k}(\xi_k, \lambda_k) = \frac{1}{2} (\augz_k - \diagg{\augH} \xi_k)^\top \diagg{\augS}_k^{-1} (\augz_k - \diagg{\augH} \xi_k) + \lambda^\top_k {\mathbb L} \xi_k$ and compute its  gradients with respect to $\xi_k$ and $\lambda_k$ as  
\begin{align}
\label{eq:gradient_estimate}
\nabla_{\xi} \mathcal{L}_{\mathsf{est},k}(\xi_k, \lambda_k) &= -\diagg{\augH}^{\top} \diagg{\augS}_k^{-1}(\augz_k - \diagg{\augH} \xi_k) + {\mathbb L} \lambda_k \\
\label{eq:gradient_estimate_dual}
\nabla_{\lambda} \mathcal{L}_{\mathsf{est},k}(\xi_k, \lambda_k) &= {\mathbb L} \xi_k.
\end{align}
From the optimality condition for ($\xi_k^*$, $\lambda_k^*$), we have the saddle point equation (KKT conditions) given by
\begin{align}\label{eq:SPE}
\begin{bmatrix} -\diagg{\augH}^\top \diagg{\augS}_k^{-1} \diagg{\augH} & -{\mathbb L} \\ {\mathbb L} & {{0}} \end{bmatrix}
\begin{bmatrix} \xi_k^* \\ \lambda_k^* \end{bmatrix} = \begin{bmatrix} -\diagg{\augH}^\top \diagg{\augS}_k^{-1} \augz_k\\ 0 \end{bmatrix}
\end{align}
where $\xi_k^* := [\xi_{1,k}^*; \cdots; \xi_{N,k}^*]$ and $\lambda_k^* := [\lambda_{1,k}^*; \cdots; \lambda_{N,k}^*]$.

\begin{lem}\label{lem:opt_pt_est} 
Suppose that $P_{i, k|k-1}$ is symmetric positive definite for all $i \in \mathcal N$. Then, the solutions to DKF problem are given by $(\xi_k^*, \lambda_k^*)=( {\mathbb 1}_N \xi_k^\dagger,  {\mathbb 1}_N \tilde \lambda_k +  \bar \lambda_k)$ where
\begin{align}\label{eq:xi_k_dagger}
\xi_k^\dagger = \mathcal{H}_k^{-1} {\mathbb 1}_N^\top \diagg{\augH}^\top \diagg{\augS}_k^{-1} \augz_k,
\end{align}
$\bar \lambda_k \in \mathbb R^n$ is a vector that is uniquely determined in terms of $W$, $\tilde{\Lambda}$, and $\xi_k^\dagger$, and $\tilde \lambda_k \in \mathbb R^n$ is an arbitrary vector. 
\end{lem}
\vspace{-0.5cm}
\begin{pf} We refer the reader to the paper \cite{Ryu2019CDC}. \hfill  $\square$
\end{pf}
\vspace{-0.5cm}
The covariance correction step can also be formulated as an optimization problem. It is formulated under the framework of information filtering \cite{Thrun2005Book}. Let $\Omega_k := P_k^{-1}$ and $P_{k}^{-1} \hat x_{k}$ be the information matrix  and information vector, respectively. We also define $\Omega_{k|k-1} = P_{k|k-1}^{-1}$.

At each time $k$, the information matrix is updated in two steps; prediction and correction, namely
\begin{subequations}
\begin{align}
\label{eq:info_mat_rule_pre}
    \Omega_{k|k-1} &= (F \Omega_{k-1}^{-1} F^\top + Q)^{-1}\\
\label{eq:info_mat_rule_CIF}
    \Omega_k &= \Omega_{k|k-1} + H^\top \diagg{R}^{-1} H.
\end{align}
\end{subequations} 
\ifnote{\begin{equation}\label{eq:info_mat_rule_pre}
\Omega_{k|k-1} = (F \Omega_{k-1}^{-1} F^\top + Q)^{-1},
\end{equation}
and in the correction step, it is updated as
\begin{equation}\label{eq:info_mat_rule_CIF}
    \Omega_k = \Omega_{k|k-1} + H^\top \diagg{R}^{-1} H.
\end{equation}}
It is noted that if the information matrix is updated by  \eqref{eq:info_mat_rule_pre} and \eqref{eq:info_mat_rule_CIF}, then $\Omega_{k|k-1}$ converges to $P^{*-1}$ where $P^*$ is a unique positive definite solution to the  discrete-time algebraic Riccati equation \cite{Anderson2012OptimalFiltering} given by 
\begin{equation}\label{eq:ARE}
P^* \!= \!F P^* F^\top \! - \!F P^* H^\top (H P^* H^\top + \diagg{R})^{-1} H P^* F^\top \!+ Q, 
\end{equation}
and the limit of $\Omega_{k|k-1}$ is the same no matter what the initial condition $\Omega_0$ is chosen as long as $\Omega_0> 0$.

Thanks to the convergence property, it is expected that $\Omega_{i,k|k-1}$ will converge to $P^{*-1}$ provided that the estimators can compute the global information rate matrix $H^\top \bar R^{-1} H$ in a distributed way; the estimators need not choose the same initial condition for $\Omega_{i,k}$. In fact, this idea is widely used, see, e.g., \cite{Olfati2005CDC, Olfati2007CDC, Olfati2009CDC, KamgarpourTomlin2008CDC}. Accordingly, we formulate a consensus optimization problem as
\begin{equation*}\tag{P.2}\label{p:cov}
\begin{split}
    \underset{\theta_{1},\dots, \theta_{N}}{\minimize}& \quad \frac{1}{2} \sum^N_{i=1} \| N \omega^\delta_i - \theta_i \|^2\\
    \st& \quad \theta_1 = \cdots = \theta_N
\end{split}
\end{equation*}
where $\omega^\delta_i = \VEC{H_i^\top R_i^{-1} H_i} \in \mathbb R^{n_{\mathsf{cov}}}$, $\theta_i \in \mathbb R^{n_{\mathsf{cov}}}$ is estimator $i$'s decision variable, and $n_{\mathsf{cov}} = n(n+1)/2$.

To solve \eqref{p:cov}, we construct the Lagrangian as $\mathcal L_{\mathsf{cov}} (\theta, \upsilon) = \frac{1}{2} (N \omega^\delta - \theta)^\top (N \omega^\delta - \theta) + \upsilon^\top \hat{\mathbb{L}} \theta$, 
where $\upsilon \in \mathbb R^{N n_{\mathsf{cov}}}$ is the Lagrange multiplier, $\omega^\delta = [\omega^\delta_1; \cdots; \omega^\delta_N]$, $\theta = [\theta_1;\cdots;\theta_N]$, and $\hat{\mathbb L} = L \otimes I_{n_{\mathsf{cov}}}$, and this leads to the following result, where we use the notation $\hat {\mathbb 1}_N$, $\hat {\mathbb I}_N$, $\hat {\mathbb U}$, $\hat {\mathbb W}$, $\hat{\mathbb \Lambda}$, and $\hat {\tilde {\mathbb \Lambda}}$ defined similarly to $\hat {\mathbb L}$.

\begin{lem}\label{lem:opt_pt_cov}
The solutions to \eqref{p:cov} are parameterized as $(\theta^*, \upsilon^*) = (\hat{\mathbb 1}_N \theta^\dagger, \hat{\mathbb 1}_N \tilde \upsilon + \bar \upsilon)$ where $\theta^\dagger = {\emph{vech}}(\Theta^\dagger)$, $\Theta^\dagger:= \sum_{i=1}^N (H_i^\top R_i^{-1} H_i)=H^\top \bar R^{-1}H$,
$\bar \upsilon = \hat{\mathbb W} \hat{\tilde{\mathbb{\Lambda}}}^{-1} \hat{\mathbb W}^\top \left(N \omega^\delta - \hat{\mathbb 1}_N \theta^\dagger \right)$, and $\tilde \upsilon \in \mathbb R^{n_{\mathsf{cov}}}$ is arbitrary. 
\end{lem}
\vspace{-0.5cm}
\begin{pf}
Similar to Lemma \ref{lem:opt_pt_est} and thus omitted. 
\hfill $\square$
\end{pf}
\vspace{-0.5cm}

\section{Dual Ascent Distributed Kalman Filtering}\label{sec:DA_DKF}\vspace{-0.2cm}
In this section, we propose a new DKF algorithm employing a distributed optimization method. The proposed DKF algorithm consists of two steps, prediction and correction, as CKF does. In the prediction step, each estimator predicts the estimate and covariance locally.
The correction step, which is our main concern, is the process of finding the minimizers of \eqref{p:estimate} and \eqref{p:cov} in a distributed fashion. For this, various distributed optimization algorithms can be employed allowing each estimator to reach the minimizers. We call this step distributed correction. 
In this paper, we employ the dual ascent method \cite{Boyd+2011FTML} for the distributed correction step, which is a well-known convex optimization method. The update of the dual variable is performed by using the gradient ascent and the primal variable is updated by finding the minimizer of the local Lagrangian with the updated dual variable. In each correction step, additional iterations can be conducted so that the primal variable of each estimator converges to the minimizer with small error. Such iteration is called subiteration and is indexed by the subscript $l$ hereafter.

For \eqref{p:estimate}, the dual ascent based update rule is given by 
\begin{subequations}\label{eq:est_rule}
\begin{align}
\label{eq:est_rule_dual}
    \lambda_{k,l+1} & = \lambda_{k,l} + \bar A_\lambda
    K_k^{\mathsf{dual}} {\mathbb L} \xi_{k,l}\\
\label{eq:est_rule_primal}
    \xi_{k,l+1} &= K_k^{\mathsf{cons}} (\diagg{\augH}^{\top} \diagg{\augS}_k^{-1} \augz_k - {\mathbb L} \lambda_{k,l+1})
\\
K_k^{\mathsf{cons}}&=(\diagg{\augH}^\top \diagg{\augS}_k^{-1} \diagg{\augH})^{-1}, \ 
K_k^{\mathsf{dual}}= \diag\{K_{1,k}^{\mathsf{dual}}, \cdots, K_{N,k}^{\mathsf{dual}}\} \nonumber\\
K_{i,k}^{\mathsf{dual}} &= \textstyle{\frac{1}{\|N P_{i,k|k-1}\| + \epsilon_i}I_n} \nonumber 
\end{align}
\end{subequations}
where $\bar A_\lambda = A_\lambda \otimes I_n$, $A_\lambda = \diag\{\alpha_{\lambda,1}, \dots, \alpha_{\lambda,N}\}$, $\alpha_{\lambda, i} > 0, i=1,\dots,N$ is the update gain and $\epsilon_i$ is an arbitrary positive scalar. When $l$ reaches $l^*$ that is the number of subiterations, the distributed correction step stops and we update $\hat x_{i,k} = \xi_{i,k,l^*}$.

One advantage of using the dual ascent for consensus optimization is that the resulting algorithm has a distributed form owing to the structure of 
$\nabla_{\xi}\mathcal{L}_{\mathsf{est},k}(\xi_k, \lambda_k)$ and $
\!\nabla_{\lambda} \mathcal{L}_{\mathsf{est},k}(\xi_k, \lambda_k)$; see \eqref{eq:gradient_estimate} and \eqref{eq:gradient_estimate_dual}. In addition, $(\xi_{k,l}, \lambda_{k,l})$ satisfies the dual feasibility equation for any $k$ and $l$, i.e., $-\diagg{\augH}^{\top} \diagg{\augS}_k^{-1}(\augz_k - \diagg{\augH} \xi_{k,l}) + {\mathbb L} \lambda_{k,l} = 0, \forall k,l$ \cite{Boyd+2011FTML}. 

It is emphasized that the update gain for the dual variable update \eqref{eq:est_rule_dual}  is chosen as $\bar A_\lambda K_k^{\mathsf{dual}}$ rather than a scalar $\alpha_{\lambda}$. Using a scalar update gain is typical in dual ascent approach and it is also the case with the preliminary version of this paper \cite{Ryu2019CDC}. In \cite{Ryu2019CDC}, the update gain is chosen using the maximum norm of local covariance, which is not easy to obtain in advance.
As will be seen in the stability proof (Lemma \ref{lem:est_rule}), the matrix update gain $\bar A_\lambda K_k^{\mathsf{dual}}$ relaxes the dependence so that the new gain with $\alpha_{\lambda,i}\le 2/\bar \sigma^2$ ($\bar \sigma$ is the upper bound of the maximum eigenvalue of $L$) ensures the stability of the proposed approach.
The rationale behind this choice is that with this new gain matrix, the dynamics of $\xi_{k,l}$ has $n$ simple eigenvalues at $1$ while the other eigenvalues are stable for sufficiently large $k$, independently of local covariance.

\begin{algorithm}[t]
\caption{DA-DKF}\label{algo:DA_DKF}
\begin{algorithmic}[1]
\State {Initialization}: Take arbitrary $\hat x_{i,0}$, $P_{i,0} > 0$, and $\epsilon_i > 0$. Set $\theta_{i,0,l^*} = \VEC{H_i^\top R_i^{-1} H_i}$, $\upsilon_{i,0,l^*} = 0$, $k=1$.
\State {\bf{repeat}}
\State \quad {\it{// Local prediction}}
\State \quad $\hat x_{i,k|k-1} = F \hat x_{i,k-1}$, $P_{i, k|k-1} = F P_{i, k-1} F^\top + Q$
\vspace{0.2cm}
\State \quad {\it{// Distributed correction}} 
\State \quad $\xi_{i,k,0} = \hat x_{i,k|k-1}, \lambda_{i,k,0} = 0$
\State \quad $\theta_{i,k,0} = \theta_{i,k-1,l^*}, \upsilon_{i,k,0} = \upsilon_{i,k-1,l^*}$
\State \quad {\bf{while}} $l= 0,\dots, l^*-1$, {\bf{do}}
\State \qquad {\it{// Distributed estimate update}} \eqref{eq:est_rule}
\State \qquad $\lambda_{i,k,l+1} = \lambda_{i,k,l} + \alpha_{\lambda,i} K_{i,k}^{\mathsf{dual}} \sum_{j \in \mathcal N_j} a_{ij}(\xi_{i,k,l} - \xi_{j,k,l})$
\State \qquad $\xi_{i,k,l+1} = \hat x_{i,k|k-1} + K_{i,k}^{\mathsf{inno}}(y_{i,k} - H_i \hat x_{i,k|k-1})$
\Statex \qquad \qquad \qquad $- K_{i,k}^{\mathsf{cons}} \sum_{j \in \mathcal{N}_i} a_{ij} (\lambda_{i,k,l+1} - \lambda_{j,k,l+1})$
\vspace{0.2cm}
\State \qquad {\it{// Distributed information rate update}} (\ref{eq:cov_rule})
\State \qquad $\upsilon_{i,k,l+1} = \upsilon_{i,k,l} + \alpha_{\upsilon,i} \sum_{j \in \mathcal N_i} a_{ij}(\theta_{i,k,l} - \theta_{j,k,l})$
\State \qquad $\theta_{i,k,l+1} = N \omega^\delta_i - \sum_{j \in \mathcal N_i} a_{ij}(\upsilon_{i,k,l+1} - \upsilon_{j,k,l+1})$
\State \quad {\bf{end}}
\State \quad $\hat x_{i,k} = \xi_{i,k,l^*}$, $P_{i, k} = (P_{i,k|k-1}^{-1} + \Theta_{i,k,l^*})^{-1}$
\State \quad $k=k+1$
\State \bf{until} $k=\infty$.
\end{algorithmic}
\hrule
\vspace{0.1cm}
\footnotesize{
$K_{i,k}^{\mathsf{inno}} = (H_i^{\top} R_i^{{-1}} H_i + \frac{1}{N} P_{i, k|k-1}^{{-1}})^{-1} H_i^{\top} R_i^{-1}$

$K_{i,k}^{\mathsf{cons}} \!=\! (H_i^\top R_i^{-1} H_i \!+\! \frac{1}{N} P_{i, k|k-1}^{-1})^{-1}$, $K_{i,k}^{\mathsf{dual}} = \textstyle{\frac{1}{\|N P_{i,k|k-1}\| + \epsilon_i}I_n}$


$\omega^\delta_i = \VEC{H_i^\top R_i^{-1} H_i}$, \quad $\Theta_{i,k,l^*} = \invVEC{\theta_{i,k,l^*}}$
}
\end{algorithm} 

Similarly to the problem \eqref{p:estimate}, the distributed correction algorithm for the problem \eqref{p:cov} can be obtained as
\begin{subequations}\label{eq:cov_rule}
\begin{align}
    \label{eq:cov_rule_dual}
    \upsilon_{k,l+1} & = \upsilon_{k,l} + \bar A_\upsilon \hat{\mathbb L} \theta_{k,l} \\
    \label{eq:cov_rule_primal}
    \theta_{k,l+1} &= N \omega^\delta - \hat{\mathbb L} \upsilon_{k,l+1} 
\end{align}
\end{subequations}
where $\bar A_\upsilon$ is defined similarly to $\bar A_\lambda$ and $\alpha_{\upsilon,i} > 0$ is the update gain.  
At the end of the distributed correction, we have $P_{i,k} = (P^{-1}_{i,k|k-1} + \Theta_{i,k,l^*})^{-1}$.
Updating the local covariance matrix with the exchanged information rate matrix is widely used in existing DKF algorithms \cite{Olfati2007CDC, Battistelli2015TAC, Marelli+2021Aut, Qian+2022Aut}. See also \cite{Zorzi2019TCNS} for the case with system uncertainty. 


Using \eqref{eq:est_rule} and \eqref{eq:cov_rule}, we propose DA-DKF (dual-ascent based distributed Kalman filtering) described in Algorithm \ref{algo:DA_DKF}. The design parameters are $\alpha_{\lambda,i}$, $\alpha_{\upsilon,i}$, $\epsilon_i$, and $l^*$. Among these parameters, $\alpha_{\lambda,i}$ and $\alpha_{\upsilon,i}$ should be selected so that the primal variable of each problem converges to the minimizer, as $l$ increases. Meanwhile, a small number of subiterations is preferred in practice in order to reduce communication and computation load. 

In the next section, we provide a sufficient condition that ensures 
the stability of DA-DKF, i.e., 
\begin{equation}\label{eq:goal}
\begin{split}
 \!\lim_{k \rightarrow \infty} \mathbb E\{x_k \!-\! \xi_{i,k,l^*}\} &= 0, ~ \lim_{k \rightarrow \infty} \|P^* \!-\! P_{i,k|k-1}\| = 0.\!
\end{split}
\end{equation}
where $P^*$ is the unique positive definite solution to \eqref{eq:ARE}.
In fact, \eqref{eq:goal} is the key stability results in the case of CKF \cite{Jazwinski2007Book, Kamen1999Book}, which implies that CKF is unbiased and converges to the steady-state CKF asymptotically. Therefore, if \eqref{eq:goal} holds true, we state that DA-DKF asymptotically recovers the performance of CKF.

\section{Stability Analysis}\label{sec:stability}
We start the stability analysis by noting that in Algorithm 1 the update rules for the estimate and the covariance are of cascade type and the latter is autonomous. 
Based on this fact, we first state results on the local covariances, which include a sufficient condition on the update gain for \eqref{p:cov} and the boundedness of the local covariances (Lemmas \ref{lem:cov_rule} and \ref{lem:cov_bound}). 
The convergence of local covariances is proved in Theorem \ref{thm:cov_convergence}. Once the boundedness of covariance is guaranteed, we establish a sufficient condition for the update gain $\alpha_{\lambda,i}$ that guarantees asymptotic convergence of the residual $e^\xi_{k,l} := \xi_k^* - \xi_{k,l}$ as $l$ increases. Then, we derive the dynamics of the expectation of estimate error, denoted by $\mathbb E \{e_k\}$ where $e_k := [e_k^\dagger; e_k^\xi]$, $e_k^\dagger := x_k - \xi_k^\dagger$, and $e_k^\xi := \xi_k^* - \xi_k$. Subsequently, convergence of $\mathbb E \{e_k\}$ is proved in Theorem \ref{thm:est_stability}.


\begin{lem}\label{lem:cov_rule}
Suppose that Assumption \ref{as:network} holds true and let $l^*$ be any positive integer. Then, the sequence $\{\Theta_{i,k,l^*} := {\emph{vech}}^{-1}{(\theta_{i,k,l^*}})\}$ generated by Algorithm \ref{algo:DA_DKF} converges to the global information rate matrix $H^\top \diagg{R}^{-1} H$, as $k$ goes to infinity, if the update gain $\alpha_{\upsilon, i}$ is chosen such that 
\begin{equation}\label{eq:alpha_ups}
    0 < \alpha_{\upsilon,i} < 2/{\bar \sigma^2}. 
\end{equation}
Moreover, if $\alpha_{\upsilon, i}$ satisfies \eqref{eq:alpha_ups}, then for any $0<\kappa<1$ there exists $\bar k>0$ such that \begin{equation}\label{eq:Theta_ulbound}
     0 \le (1 - \kappa) \Theta^\dagger \leq \Theta_{i,k,l^*} \leq (1+\kappa) \Theta^\dagger, \quad \forall k \geq \bar k.
\end{equation}
\end{lem}
\vspace{-0.7cm}
\begin{pf}
See Appendix \ref{appdx:cov_rule}. \hfill  $\square$
\end{pf}
\vspace{-0.5cm}

\begin{rem}\label{rmk:cov}
Since $P_{i,k}$ is an estimate of covariance, it is meaningful when $P_{i,k}\ge 0$. In fact, in the case of CKF, the positive definiteness of $P_k, k\ge 1$, is guaranteed whenever $Q$ is positive definite. Unfortunately, the positive definiteness of $P_{i,k}$ generated by DA-DKF may not be preserved or may not even be well defined if $\Theta_{i,k,l^*}$ is not positive semidefinite since $P_{i,k}$ is computed by $P_{i,k}=(P_{i,k|k-1}^{-1}+\Theta_{i,k,l^*})^{-1}$ (line 16 of Algorithm \ref{algo:DA_DKF}).

In fact, the positive definiteness can be guaranteed by making $\Theta_{i,k,l^*}$ positive semidefinite. This is because the iteration of $\Theta_{i,k,l}$ (lines 13 and 14 of the algorithm) is done autonomously and independently of $P_{i,k}$. Based on this observation and the result given in Lemma  \ref{lem:cov_rule}, we propose two simple ways to keep $P_{i,k}$ to be positive definite for all $k \geq 1$; i) using sufficiently large $l^*$ and ii) exception handling when $\Theta_{i,k,l^*}$ is not positive semidefinite. 

For the former case, use a sufficiently large $\bar l^*$ such that $\Theta_{i,k,l^*} \geq 0$ for any $k \geq 1$, $l^* \geq \bar l^*$. Note that the existence of $\bar l^*$ is clear from Lemma \ref{lem:cov_rule}. 
One example of exception handling would be to use a projected matrix of $\Theta_{i,k,l^*}$ onto a set of positive semidefinite matrices, denoted by $ \text{\rm Proj}_{\ge 0}(\Theta_{i,k,l^*})$, to  compute $P_{i,k}$, namely, modify line 16 of Algorithm 
\ref{algo:DA_DKF} as $P_{i,k} = (P_{i,k|k-1}^{-1} + \text{\rm Proj}_{\ge 0}(\Theta_{i,k,l^*}))^{-1}$.
Obtaining $\text{\rm Proj}_{\ge 0}(\Theta_{i,k,l^*})$ can be accomplished by formulating a semidefinite programming and solving it \cite{Boyd+2011FTML}. A simple and effective way to find that matrix, using an image of $P_{i,k|k-1}$, is introduced in \cite{YiZorzi2021TAC}.
\end{rem}

\begin{lem}\label{lem:cov_bound} 
Consider Algorithm \ref{algo:DA_DKF} (or the modified one discussed in Remark \ref{rmk:cov}) and suppose that Assumptions \ref{as:collecObs} and \ref{as:network} hold true. Then, if the update gain $\alpha_{\upsilon,i}$ is chosen such that \eqref{eq:alpha_ups} is satisfied, there exist symmetric positive definite matrices $\underline P > 0$ and $\overline P < \infty$ such that $\underline P \leq P_{i,k} \leq \overline P$ and $\underline P \leq P_{i,k|k-1} \leq \overline P$ for all $k \geq 1$ and $i \in \mathcal N$.
\end{lem}
\vspace{-0.5cm}
\begin{pf} See Appendix \ref{appdx:cov_bound}.  \hfill  $\square$
\end{pf}
\vspace{-0.5cm}
In CKF, it is known that the covariance converges to $P^*>0$ that is a unique positive definite solution of the discrete-time algebraic Riccati equation 
\begin{equation}\label{eq:DARE_CKF}
    P = F \left \{P - P H^\top (H^\top P H + \diagg{R})^{-1} H P \right \} F^\top + Q
\end{equation}
for any initial covariance $P_{0} \geq 0$ \cite{ Jazwinski2007Book}. In the following theorem, we state the asymptotic performance recovery of DKF to CKF in terms of covariance. 

\begin{thm}\label{thm:cov_convergence}
Suppose that Assumptions \ref{as:collecObs} and \ref{as:network} hold true, and the update gain $\alpha_{\upsilon,i}$ is chosen such that $0 < \alpha_{\upsilon,i} <2/\bar \sigma^2$. Then, the local covariance matrix $P_{i,k|k-1}$ generated by Algorithm \ref{algo:DA_DKF} converges to the unique positive definite solution of \eqref{eq:DARE_CKF}, i.e., that of CKF. 
\end{thm}
\vspace{-0.5cm}
\begin{pf}
From the covariance prediction rule given by $P_{i,k+1|k} = F P_{i,k} F^\top + Q$, we have 
\begin{equation}\label{eq:P_i_iteration}
\begin{split}
    P_{i,k+1|k} &= F (P_{i,k|k-1}^{-1} + \Theta_{i,k,l^*})^{-1} F^\top + Q\\
    &= F (P_{i,k|k-1}^{-1} + \Theta^\dagger )^{-1} F^\top + Q + M_{i,k,l^*}
    \end{split}
\end{equation}
where $M_{i,k,l^*} = F (P_{i,k|k-1}^{-1} + \Theta_{i,k,l^*})^{-1} F^\top - F (P_{i,k|k-1}^{-1} + \Theta^\dagger )^{-1} F^\top$. By Lemma \ref{lem:cov_rule}, it holds that $\Theta_{i,k,l^*} \geq 0$ for all $k \geq \bar k$, $i \in \mathcal N$. Then, we can rewrite $M_{i,k,l^*}$ as 
\begin{equation*}
    M_{i,k,l^*} = F (P_{i,k|k-1}^{-1} + \Theta_{i,k,l^*})^{-1} e_{i,k,l^*}^\Theta (P_{i,k|k-1}^{-1} + \Theta^\dagger )^{-1} F^\top 
\end{equation*}
where $e_{i,k,l^*}^\Theta := \Theta^\dagger - \Theta_{i,k,l^*}$, and this leads to $\|M_{i,k,l^*}\| \leq {\bar f}^2 {\bar p}^2 \| e_{i,k,l^*}^\Theta \|$
where $\bar p$ is a positive scalar such that $\|\overline P\| \leq \bar p$. Since $\lim_{k \rightarrow \infty} e_{i,k,l^*}^\Theta = 0$ by Lemma \ref{lem:cov_rule}, we have
\begin{equation}\label{eq:lim_M_ik}
    \lim_{k \rightarrow \infty} \| M_{i,k,l^*} \| = 0.
\end{equation}
Let $\tilde P_{i,k} = P^* - P_{i,k|k-1}$ where $P^*>0$ is the unique solution of \eqref{eq:DARE_CKF}. By the matrix inversion lemma, one has
\begin{equation}\label{eq:P*}
    P^* = F(P^{*-1} + \Theta^\dagger)^{-1} F^\top + Q.
\end{equation}
Then, we have from \eqref{eq:P_i_iteration} and \eqref{eq:P*} that
\begin{align}\label{eq:tilde_P}
    &\tilde P_{i,k+1} = P^* - P_{i,k+1|k} \nonumber \\
    &= F \left( (P^{*-1} + \Theta^\dagger)^{-1} - (P_{i,k|k-1}^{-1} + \Theta^\dagger )^{-1} \right) F^\top - M_{i,k,l^*} \nonumber\\
    &= F (P^{*-1} + \Theta^\dagger)^{-1} P^{*-1} \tilde P_{i,k} P_{i,k|k-1}^{-1} (P_{i,k|k-1}^{-1} + \Theta^\dagger )^{-1} F^\top \nonumber\\
    & \quad - M_{i,k,l^*}.
\end{align}
Noting that $\Theta^\dagger = H^\top \bar R^{-1} H$, one can derive
\begin{equation}\label{eq:Phi*_Phi_ik}
\begin{split}
    F (P^{*-1} + \Theta^\dagger)^{-1} P^{*-1} &= F(I_n - K^* H)\\
    F (P_{i,k|k-1}^{-1} + \Theta^\dagger)^{-1} P_{i,k|k-1}^{-1} &= F(I_n - K_{i,k}H)
\end{split}
\end{equation}
where $K^* = P^* H^\top (\diagg{R}+ HP^*H^\top)^{-1}$ and $K_{i,k} = P_{i,k|k-1} H^\top (\diagg{R}+ H P_{i,k|k-1} H^\top)^{-1}$ are the Kalman gains in the steady-state and at time $k$, respectively. Define $\Phi^* = F(I_n - K^* H)$ and $\Phi_{i,k} = F(I_n - K_{i,k}H)$. Then, from \eqref{eq:tilde_P} and \eqref{eq:Phi*_Phi_ik}, we have
\begin{equation}\label{eq:tilde_P_Lyap}
    \tilde P_{i,k+1} = \Phi^* \tilde P_{i,k} \Phi_{i,k}^\top - M_{i,k,l^*}.
\end{equation}
We now investigate $\Phi^*$ and $\Phi_{i,k}$. Since $(F, \sqrt{Q})$ is controllable and $(F, H)$ is observable, by Lemma D.2 in \cite{Kamen1999Book}, $\Phi^*$ is Schur stable and there exists  $a^*>0$ such that 
\begin{equation}\label{eq:a_star}
    \|\Phi^{*}\| \leq a^* < 1.  
\end{equation}
Regarding $\Phi_{i,k}$, we rewrite \eqref{eq:P_i_iteration} as 
\begin{equation}\label{eq:P_ik_Lyapunov1}
\begin{split}
  \!\!\!\!  P_{i,k+1|k} &\!= \Phi_{i,k} P_{i,k|k-1} \Phi_{i,k}^\top + F P_{i,k|k-1} H^\top K_{i,k}^\top F^\top \\
    &\! \!\!\!\!\!\!\!\!\quad - F K_{i,k} H P_{i,k|k-1} H^\top K_{i,k}^\top F^\top + Q + M_{i,k,l^*}.\!\!\!\!
\end{split}
\end{equation}
Recalling that $K_{i,k} = P_{i,k|k-1} H^\top (\diagg{R}+ H P_{i,k|k-1} H^\top)^{-1}$, we have $K_{i,k} H P_{i,k|k-1} H^\top K_{i,k}^\top = P_{i,k|k-1} H^\top K_{i,k}^\top - K_{i,k} \bar R K_{i,k}^\top$.  
Then, applying this identity to \eqref{eq:P_ik_Lyapunov1} yields
\begin{equation}\label{eq:P_ik_Lyapunov2}
    P_{i,k+1|k} = \Phi_{i,k} P_{i,k|k-1} \Phi_{i,k}^\top + K_{i,k} \bar R K_{i,k}^\top + Q + M_{i,k,l^*}.
\end{equation}
From \eqref{eq:lim_M_ik}, there exists a sufficiently large $k^{*}_{{\rm cov}}$ such that $Q + M_{i,k,l^*} > 0$, $\forall k \geq k^{*}_{{\rm cov}}$. Then, it follows from \eqref{eq:P_ik_Lyapunov2} that
    $\Phi_{i,k:j} P_{i,j|j-1} \Phi_{i,k:j}^\top \leq \Phi_{i,k} P_{i,k|k-1} \Phi_{i,k}^\top \leq P_{i,k+1|k}$ 
where $\Phi_{i,k:k-j} = \Phi_{i,k} \Phi_{i,k-1} \dots \Phi_{i,k-j}$  for any $j$ such that $k^{*}_{{\rm cov}} \leq j \leq k$ (clearly, $\Phi_{i,k:k} = \Phi_{i,k}$). Since $P_{i,k+1|k} \leq \overline{P}$ by Lemma \ref{lem:cov_bound}, there exists $a_i>0$ such that for an arbitrary $j$ such that $k^{*}_{{\rm cov}} \leq j \leq k$ (clearly, $\Phi_{i,k:k} = \Phi_{i,k}$). Since it holds that $P_{i,k+1|k} \leq \overline{P}$ by Lemma \ref{lem:cov_bound}, there exists $a_i>0$ such that 
\begin{equation}\label{eq:a_i}
    \|\Phi_{i,k:j}\| \leq a_i, \quad \forall k \geq j \geq k^{*}_{{\rm cov}}.
\end{equation}
Then, we obtain from \eqref{eq:tilde_P_Lyap}, \eqref{eq:a_star}, and \eqref{eq:a_i} that, for any $k \geq k^{*}_{{\rm cov}}$, $\|\tilde P_{i,k+1}\|  \leq a_i a^* \|\tilde P_{i,k}\| + \| M_{i,k,l^*} \|  \leq a_i (a^*)^{(k - k^{*}_{{\rm cov}})} \|\tilde P_{i,k^*}\| + a_i \sum_{j = k^{*}_{{\rm cov}}}^{k} (a^*)^{(k - j)} \| M_{i,j,l^*} \|$. 
Finally, applying \eqref{eq:a_star} and \eqref{eq:lim_M_ik} to the preceding inequality yields $\lim_{k \rightarrow \infty} \| P^* - P_{i,k+1|k} \| = 0, \   \forall i \in \mathcal N$. \hfill $\square$
\end{pf}
\vspace{-0.3cm}
Now we present a sufficient condition on $\alpha_{\lambda,i}$ that guarantees the convergence of the primal variable of \eqref{p:estimate}.



\begin{lem}\label{lem:est_rule}
Suppose that Assumption \ref{as:network} holds true. Consider Algorithm \ref{algo:DA_DKF} with the update gains $\alpha_{\upsilon,i}$ and $\alpha_{\lambda,i}$ satisfying
\begin{equation}\label{eq:alpha_upsilon_lambda}
0<\alpha_{\upsilon,i} < 2/{\bar \sigma^2}, \ 0 < \alpha_{\lambda, i} < 2/{\bar \sigma^2}.
\end{equation} 

Then, there exists $k^*\ge 1$ such that for any $k\ge k^*$,  $\xi_{k,l}$ converges to $\xi_k^*$ as $l$ goes to infinity and $\lambda_{k,l}$ is bounded for all $l$. 
Moreover, there exist $0<\mu<1$ and $k^* \ge 1$ such that $\|\tilde{\Xi}_{k}\| \leq \mu, \forall k \geq k^*$ where 
$\tilde{\Xi}_k = {\mathbb I}_{N-1} - {\mathbb W}^\top K_k^{\mathsf{cons}} {\mathbb W} \tilde{\mathbb \Lambda} {\mathbb W}^\top \bar A_\lambda K_k^{\mathsf{dual}} {\mathbb W} \tilde{\mathbb \Lambda}$.
\end{lem}
\vspace{-0.5cm}
\begin{pf}
See Appendix \ref{appdx:est_rule}. \hfill  $\square$
\end{pf}
\vspace{-0.5cm}
Lemma \ref{lem:est_rule} has established that  $\xi_{k,l}$ converges to the optimal point $\xi_k^*$ as the subiteration proceeds, i.e., $l$ tends to infinity. However, this does not imply the success of the estimation since only a finite number of subiterations are carried out at each time $k$ and the point $\xi_k^*$ varies with $k$. Hence, we need to investigate the behavior of variables under DA-DKF algorithm as $k$ increases. 

To proceed, we define the following.
\begin{align*}
        \mathsf e_k^\dagger &= \mathbb E\{x_k - \xi_k^\dagger\},
        ~~ \mathsf e_{k,l}^\xi = {\mathbb E}\{\xi^* - \xi_{k,l}\}
        ,~~ \breve{\mathsf e}_{k,l}^\xi = \mathbb U^\top {\mathsf e}_{k,l}^\xi\\
        \diagg{F} &= I_N \otimes F, ~~ \diagg{H} = \diag\{H_1, \dots, H_N\}\\
        \bar P_{k+1|k} &= \diag\{P_{1,k+1|k} , \dots,P_{N,k+1|k}\}.
\end{align*}

The following result is on the structure of the dynamics of $[{\mathsf e}_k^{\dagger}; \breve{\mathsf e}_{k,l}^\xi]$, which plays a key role in the stability proof.
\begin{lem}\label{lem:error_dyn}
Consider $\xi_{i,k,l^*}$ generated by Algorithm \ref{algo:DA_DKF}. Then, the dynamics of $\msf{e}_k := [{\mathsf e}_k^{\dagger}; \breve{\mathsf e}_{k,l}^\xi]$ is given by 
\begin{equation}\label{eq:error_dyn}
\begin{split}
    \begin{bmatrix}
    \msf{e}_{k+1}^\dagger \\
    \breve{\msf{e}}_{k+1,l^*}^\xi \\
    \end{bmatrix}
    &= 
    \begin{bmatrix}
    E^\dagger_{k+1} & E^\ddagger_{k+1} \\
    0 & \breve{E}^{\xi}_{k+1} (\breve{\Xi}_{k+1} )^{l^*} \diagg{F} 
    \end{bmatrix}
    \begin{bmatrix}
    \msf{e}_{k}^\dagger \\
    \breve{\msf{e}}_{k,l^*}^\xi \\
    \end{bmatrix}
\end{split}
\end{equation}
where $\breve{E}^\xi_{k} = {\mathbb U}^\top \big({\mathbb I}_N - {\mathbb 1}_N \mathcal{H}_{k}^{-1} {\mathbb 1}_N^\top (K_k^{\mathsf{cons}})^{-1} \big) {\mathbb U}$, $E^\dagger_{k} = \textstyle \mathcal{H}_{k}^{-1} {\mathbb 1}_N^\top \frac{1}{N} \bar P_{k|k-1}^{-1} {\mathbb 1}_N F$, $E^\ddagger_{k} = \textstyle \mathcal{H}_{k}^{-1} {\mathbb 1}_N^\top \frac{1}{N} \bar P_{k|k-1}^{-1} \diagg{F} {\mathbb U}$, and
$\breve{\Xi}_{k} = {\mathbb U}^\top \left({\mathbb I}_N  - K_k^{\mathsf{cons}}{\mathbb L} \bar A_\lambda K_k^{\mathsf{dual}} \right) {\mathbb U}$. 
Moreover, if $\alpha_{\lambda,i}$ is chosen such that $0<\alpha_{\lambda,i} < 2/\bar \sigma^2$, then there exists $k^*\ge 1$ such that $\breve{E}^{\xi}_{k} (\breve{\Xi}_{k})^{l^*}$ is Schur stable for all $k \ge  k^*$. 
\end{lem}
\vspace{-0.5cm}
\begin{pf}
See Appendix \ref{appdx:error_dyn}. \hfill  $\square$
\end{pf}
\vspace{-0.5cm}
Finally, we state the stability result on estimates.
\begin{thm}\label{thm:est_stability}
Suppose that Assumptions \ref{as:collecObs} and \ref{as:network} hold true and consider Algorithm \ref{algo:DA_DKF}. If $\alpha_{\lambda,i}$ and $\alpha_{\upsilon,i}$ are chosen such that $0 < \alpha_{\lambda,i} < 2/\bar \sigma^2$ and $0 < \alpha_{\upsilon,i} < 2/\bar \sigma^2$, then there exists $\overline l^*$ such that for any $l^* \geq \overline l^*$, it holds that
\begin{equation*}
    \lim_{k \rightarrow \infty} \mathbb E\{x_k - \xi_{i,k,l^*} \} = 0. 
\end{equation*} 
\end{thm}
\vspace{-0.7cm}
\begin{pf}
Let $V_{k}^\dagger = \msf{e}_{k}^{\dagger\top} \mathcal{H}_{k} \msf{e}_{k}^\dagger$ and compute
\begin{align}\label{eq:V_k+1^dagger_begin}
\!\!\!   V_{k+1}^\dagger &=  -\msf{e}_{k+1}^{\dagger\top} ( \mathcal{H}_{k+1} - 2\mathcal{H}_{k+1} )\msf{e}_{k+1}^\dagger \nonumber\\
&= - \msf{e}_{k+1}^{\dagger\top} {\mathbb 1}_N^\top \diagg{H}^\top \diagg{R}^{-1} \diagg{H} {\mathbb 1}_N  \msf{e}_{k+1}^\dagger\\
&\quad - \msf{e}_{k+1}^{\dagger\top} {\mathbb 1}_N^\top \frac{1}{N} \bar P_{k+1|k}^{-1} {\mathbb 1}_N  \msf{e}_{k+1}^\dagger   + 2\msf{e}_{k+1}^{\dagger\top} \mathcal{H}_{k+1}\msf{e}_{k+1}^\dagger.\!\!\!  \nonumber
\end{align}
Define $\mathcal P_{k+1|k} = \left( {\mathbb 1}_N^\top \frac{1}{N} \bar P_{k+1|k}^{-1} {\mathbb 1}_N  \right)^{-1}$ and $\breve{\mathsf e}_k^\xi=\breve{\mathsf e}_{k,l^*}^\xi$.  Recalling that $\diagg{H} = \diag\{H_1,\! \dots,\! H_N\}$ and $\bar P_{k+1|k} = \diag\{P_{1,k+1|k} ,\! \dots,\! P_{N,k+1|k}\}$, we rewrite the dynamics of $\msf{e}_{k+1}^\dagger$ in \eqref{eq:error_dyn} as 
\begin{equation}\label{eq:e_dagger_k+1_to_e_dagger_k}
\begin{split}
& \mathcal P_{k+1|k} \mathcal{H}_{k+1} \msf{e}_{k+1}^\dagger = F \msf{e}_{k}^\dagger + \breve G^\xi_{k+1} \breve{\msf{e}}_{k}^\xi
\end{split}
\end{equation}
where $\breve G^\xi_{k+1} = - \mathcal P_{k+1|k} {\mathbb 1}_N^\top \frac{1}{N} \bar P_{k+1|k}^{-1} {\mathbb U} \diagg{F}$.

With \eqref{eq:e_dagger_k+1_to_e_dagger_k}, the last term of \eqref{eq:V_k+1^dagger_begin} can be written as $2\msf{e}_{k+1}^{\dagger\top} \mathcal{H}_{k+1} \msf{e}_{k+1}^\dagger = 2\msf{e}_{k+1}^{\dagger\top} \mathcal{P}_{k+1|k}^{-1} (F \msf{e}_{k}^\dagger + \breve G^\xi_{k+1} \breve{\msf{e}}_{k}^\xi)$,
from which, \eqref{eq:V_k+1^dagger_begin} becomes 
$V_{k+1}^\dagger = - \msf{e}_{k+1}^{\dagger\top} \mathcal H_{k+1} \msf{e}_{k+1}^\dagger + 2\msf{e}_{k+1}^{\dagger\top} \mathcal P_{k+1|k}^{-1} F \msf{e}_{k}^\dagger  + 2 \msf{e}_{k+1}^{\dagger\top} \mathcal P_{k+1|k}^{-1} \breve G^\xi_{k+1} \breve{\msf{e}}_{k}^\xi$.


In addition, following \cite{Jazwinski2007Book} and \cite{Kamen1999Book}, we rewrite $\msf{e}_{k+1}^\dagger$ as 
\begin{equation}\label{eq:e_dagger_with_u}
\begin{split}
\msf{e}_{k+1}^\dagger &= F \msf{e}_{k}^\dagger + (E^\dagger_{k+1} - F) \msf{e}_{k}^\dagger + E^\ddagger_{k+1} \breve{\msf{e}}_{k}^\xi\\
&=: F \msf{e}_{k}^\dagger + u_{k+1}.
\end{split}
\end{equation}
We then add and subtract $\msf{e}_{k}^{\dagger \top} F^\top \mathcal P_{k+1|k}^{-1} F \msf{e}_{k}^\dagger$ to complete the square, namely,
\vspace{-0.1cm}
\begin{equation}\label{eq:completing_square}
\begin{split}
\!\!\!\! V_{k+1}^\dagger &\!=\! -\msf{e}_{k+1}^{\dagger\top} H^\top \bar R^{-1} H \msf{e}_{k+1}^\dagger - u_{k+1}^\top \mathcal P_{k+1|k}^{-1} u_{k+1}\\
& \quad + \msf{e}_{k}^{\dagger \top} F^\top \mathcal P_{k+1|k}^{-1} F \msf{e}_{k}^\dagger + 2 \msf{e}_{k+1}^{\dagger\top} \mathcal P_{k+1|k}^{-1} \breve G^\xi_{k+1} \breve{\msf{e}}_{k}^\xi.\!
\end{split}
\end{equation}
By the matrix inversion lemma and the continuity argument, the third term $\msf{e}_{k}^{\dagger \top} F^\top \mathcal P_{k+1|k}^{-1} F \msf{e}_{k}^\dagger$ is bounded as $\msf{e}_{k}^{\dagger \top} F^\top \mathcal P_{k+1|k}^{-1} F \msf{e}_{k}^\dagger \leq \msf{e}_{k}^{\dagger \top} \frac{1}{N} \sum_{i=1}^N P_{i,k}^{-1} \msf{e}_{k}^\dagger$. Moreover, from the fact that $P_{i,k}^{-1} = P_{i,k|k-1}^{-1} + \Theta_{i,k,l^*}$ and $\sum_{i=1}^N \Theta_{i,k,l^*} = N \sum_{i=1}^N H_i^\top R_i^{-1} H_i$ for any $k \geq 1$ and $l^* \geq 1$ (see the proof of Lemma \ref{lem:cov_rule}), we have
\vspace{-0.1cm}
\begin{align}\label{eq:V_k^dagger}
    \msf{e}_{k}^{\dagger \top} F^\top \mathcal P_{k+1|k}^{-1} F \msf{e}_{k}^\dagger
    &\leq V_k^\dagger.
\end{align}
Let $\bar f>0$ be such that $\|F\| \leq \bar f$. 
Then, from \eqref{eq:error_dyn}, we know that there exist constants $c_0$ and $c_1$, which depend on $\|\overline P\|$, $\|\underline P\|$, and $\bar f$, such that 
\vspace{-0.1cm}
\begin{equation}\label{eq:c_0,c_1_bound}
\begin{split}
&2 \msf{e}_{k+1}^{\dagger\top} \mathcal P_{k+1|k}^{-1} \breve G^\xi_{k+1} \breve{\msf{e}}_{k}^\xi \leq  c_0 \|\msf{e}_{k}^{\dagger}\| \|\breve{\msf{e}}_{k}^\xi\| +  c_1 \|\breve{\msf{e}}_{k}^\xi\|^2.
\end{split}
\end{equation}
Applying the relations \eqref{eq:V_k^dagger} and \eqref{eq:c_0,c_1_bound} to \eqref{eq:completing_square} yields
$V_{k+1}^\dagger - V_k^\dagger  \leq - \msf{e}_{k+1}^{\dagger\top} H^\top \bar R^{-1} H \msf{e}_{k+1}^\dagger - u_{k+1}^\top \mathcal P_{k+1|k}^{-1} u_{k+1}
 + c_0 \|\msf{e}_{k}^{\dagger}\| \|\breve{\msf{e}}_{k}^\xi\| +  c_1 \|\breve{\msf{e}}_{k}^\xi\|^2$,
and by Young's inequality, it follows that
\vspace{-0.1cm}
\begin{align}\label{eq:Lyap_dagger}
V_{k+1}^\dagger - V_k^\dagger & \leq - \msf{e}_{k+1}^{\dagger\top} H^\top \bar R^{-1} H \msf{e}_{k+1}^\dagger - u_{k+1}^\top \mathcal P_{k+1|k}^{-1} u_{k+1} \nonumber\\
& \quad + \frac{c_0}{2\varepsilon} \|\msf{e}_{k}^{\dagger}\|^2 + \left(\frac{c_0 \varepsilon}{2} + c_1 \right) \|\breve{\msf{e}}_{k}^\xi\|^2
\end{align}
where $\varepsilon > 0$ is a constant to be determined later.
\vspace{0.2cm}

Let us define $J_{k} = \sum_{j=0}^{n-1} \{\msf{e}_{k+1+j}^{\dagger \top} H^\top \bar R^{-1} H \msf{e}_{k+1+j}^\dagger$ $+ u_{k+1+j}^{\dagger \top} \mathcal P_{k+1+j|k+j}^{-1} u_{k+1+j}^\dagger \}$ and sum up \eqref{eq:Lyap_dagger} from $k$ to $k+n$ to have
\vspace{-0.2cm}
\begin{equation}\label{eq:Lyap_dagger_k+n_to_k}
\begin{split}
    V_{k+n}^\dagger - V_{k}^\dagger &\leq 
    -J_k + \frac{c_0}{2\varepsilon} \sum_{j=0}^{n-1} \|{\msf{e}}_{k+j}^\dagger\|^2 \\
    &\quad\qquad  + \left(\frac{c_0 \varepsilon}{2} + c_1 \right) \sum_{j=0}^{n-1} \|\breve{\msf{e}}_{k+j}^\xi\|^2.
\end{split}    
\end{equation}
We would like to bound the function $J_k$. 
Let $\augE_{k+1}^\dagger = [\msf{e}_{k+1}^\dagger; \cdots; \msf{e}_{k+n}^\dagger]$ and $\augU_{k+1} = [u_{k+1}; \cdots; u_{k+n}]$. Then, from \eqref{eq:e_dagger_with_u}, one obtains
\vspace{-0.1cm}
\begin{align*}
\augE_{k+1}^\dagger &= \begin{bmatrix} F \\ F^2 \\ \vdots \\ F^n \end{bmatrix} \msf{e}_{k}^\dagger + \begin{bmatrix}
I & & & \\ F & I & & \\ \vdots & \ddots & \ddots & \\ F^{n-1} & \cdots & F & I \end{bmatrix} \augU_{k+1}\\
& =: \augF \msf{e}_{k}^\dagger + \augG \augU_{k+1}.
\end{align*}
By letting $\augHall = I_n \otimes H$, $\augRall = I_n \otimes \diagg{R}$, and $\augP_{k+1|k} = \textstyle \diag \{\mathcal P_{k+1|k}, \dots, \mathcal P_{k+n|k+n-1}\}$, one can rewrite $J_{k}$ as $J_{k} = (\augF \msf{e}^{\dagger}_{k} + \augG \augU_{k+1})^\top \augHall^\top \augRall^{-1} \augHall (\augF \msf{e}^{\dagger}_{k} + \augG \augU_{k+1})  + \augU_{k+1}^\top \augP^{-1}_{k+1|k} \augU_{k+1} $.
Since $J_k$ is convex and quadratic with respect to $\augU_{k+1}$,
one can obtain $\augU^*_{k+1}$ minimizing $J_k$ from $\nabla_{\augU} J_{k} = 0$, i.e.,  
$\augU^*_{k+1} = -(\augG^\top \mathscr H^\top \mathscr R^{-1} \mathscr H \augG + \augP_{k+1|k}^{-1})^{-1}  \augG^\top \mathscr H^\top \augRall^{-1} \mathscr H \augF \msf{e}_{k}^\dagger$.
Then, the minimum of $J_{k}$ is given by
$J^*_k = \msf{e}_k^{\dagger \top} \mathcal O^\top (\augRall + \augHall \augG \augP_{k+1|k}^{-1} \augG^\top \augHall^\top)^{-1} \mathcal O \msf{e}_k^\dagger$ 
where $\mathcal O$ is the observability matrix of the pair $(F, H)$. Since $\mathcal O$ has full column rank and $\augRall > 0$, there exists $c^\dagger > 0$ such that
\begin{equation}\label{eq:J_bound}
    2c^\dagger \|\msf{e}_k^\dagger\|^2 \leq J^*_{k} \leq J_{k}.
\end{equation}
From \eqref{eq:error_dyn}, there exists $\bar c > 0$ such that $\|\msf{e}_{k+1}^\dagger\|^2 \leq \bar c\|\msf{e}_{k}^\dagger\|^2 + \bar c\|\breve{\msf{e}}_{k}^\xi\|^2$. Then, with  $\tilde c = \max\{1, \bar c^{n-1}\}$, 
the second term in \eqref{eq:Lyap_dagger_k+n_to_k} can be bounded as
\begin{equation}\label{eq:bound_e_k+j^dagger}
\begin{split}
    \sum_{j=1}^{n-1} \|{\msf{e}}_{k+j}^\dagger\|^2 
    & \leq (n-1) \tilde c \Big\{ \|{\msf{e}}_{k}^\dagger\|^2 + \sum_{j=1}^{n-1} \|\breve{\msf{e}}_{k+j}^\xi\|^2 \Big\}.
    \end{split}
\end{equation}

In addition, from \eqref{eq:error_dyn} and the structure of $\breve{E}^\xi_{k+1} (\breve{\Xi}_{k+1})^{l^*}$, we have
\begin{equation*}\label{eq:breve_e_k+1_xi}
\begin{split}
\!\!\!\|\breve{\msf{e}}_{k+1}^\xi\|^2 &= \breve{\msf{e}}_{k}^{\xi \top} \diagg{F}^\top (\breve \Xi_{k+1}^\top)^{l^*} \breve{E}^{\xi \top}_{k+1} \breve{E}^\xi_{k+1} (\breve \Xi_{k+1})^{l^*} \diagg{F} \breve{\msf{e}}_{k}^{\xi}\\
& = \breve{\msf{e}}_{k}^{\xi \top} \diagg{F}^\top ~\! \diag \{ 0_{n\times n}, (\tilde{\Xi}_{k+1}^\top)^{l^*} (\tilde{\Xi}_{k+1})^{l^*}\} \diagg{F} \breve{\msf{e}}_{k}^{\xi}.
\end{split}
\end{equation*}
Since there exists $ \mu > 0$ such that $\|\tilde{\Xi}_{k+1}\| \leq \mu < 1$ for all $k \geq k^*$ (see Lemma \ref{lem:est_rule}), we have  
\begin{equation}\label{eq:e^xi_norm_bound}
    \|\breve{\msf{e}}_{k+1}^\xi\|^2 \leq  2 \bar f^2 \mu^{2l^*} \|\breve{\msf{e}}_{k}^{\xi}\|^2,
\end{equation}
which results in
\begin{equation}\label{eq:sum_bar_e_k+j^xi_bound}
\begin{split}
    \sum_{j=0}^{n-1} \|\breve{\msf{e}}_{k+j}^\xi\|^2 \leq \sum_{j=0}^{n-1} \left( 2 \bar f^2 \mu^{2l^*} \right)^j \|\breve{\msf{e}}_k^\xi\|^2.
\end{split}
\end{equation}
Take $\varepsilon=(n-1)c_0\tilde c/(2c^\dagger)$ and substitute \eqref{eq:J_bound}, \eqref{eq:bound_e_k+j^dagger} and \eqref{eq:sum_bar_e_k+j^xi_bound} into \eqref{eq:Lyap_dagger_k+n_to_k}. Then, one has
\begin{equation}\label{eq:Lyap_dagger_n_step}
\begin{split}
  \!\!  &V_{k+n}^\dagger - V_{k}^\dagger \leq -c^\dagger\|\msf{e}_k^\dagger\|^2 + c_2  \sum_{j=0}^{n-1} \left(2 \bar f^2 \mu^{2l^*} \right)^j \|\breve{\msf{e}}_k^\xi\|^2\!
\end{split}
\end{equation}
where $c_2= \frac{(n-1)c_0  \tilde c}{2\varepsilon} + \frac{c_0 \varepsilon}{2} +  c_1$. 

Meanwhile, let $V_{k}^\xi = \breve{\msf{e}}_{k}^{\xi \top} \breve{\msf{e}}_{k}^\xi$. Then, from \eqref{eq:e^xi_norm_bound}, one has  
$V^\xi_{k+1} \leq  2 \bar f^2 \mu^{2l^*} \|\breve{\msf{e}}_{k}^{\xi}\|^2$,
and this yields $V^\xi_{k+n}  \leq  2 \bar f^2 \mu^{2l^*} \|\breve{\msf{e}}_{k+n-1}^{\xi}\|^2  \leq \left( 2 \bar f^2 \mu^{2l^*} \right)^n \|\breve{\msf{e}}_k^{\xi}\|^2$.
Hence,
\begin{equation}\label{eq:Lyap_xi_n_step}
    V^\xi_{k+n} - V^\xi_{k} \leq - \left(1 - \left(2 \bar f^2 \mu^{2l^*}\right)^n \right) \|\breve{\msf{e}}_{k}^{\xi}\|^2.
\end{equation}
We now consider a Lyapunov function given by $V_k = V_k^\dagger + \gamma V_k^\xi$ where $\gamma$ is a positive scalar to be determined later. Then, from \eqref{eq:Lyap_dagger_n_step} and \eqref{eq:Lyap_xi_n_step}, we have 
$V_{k+n} - V_k \leq -c^\dagger\|\msf{e}_k^\dagger\|^2  + \left(c_2-\gamma(1-2\bar f^2 \mu^{2l^*}) \right) \sum_{j=0}^{n-1} \left(2 \bar f^2 \mu^{2l^*} \right)^j \|\breve{\msf{e}}_k^\xi\|^2$ 
where the identity  $\left(1 - 2 \bar f^2 \mu^{2l^*} \right) \sum_{j=0}^{n-1} \left(2 \bar f^2 \mu^{2l^*} \right)^j = 1 - \left(2 \bar f^2 \mu^{2l^*}\right)^n$ is used. Choose $l^*$ such that $2 \bar f^2 \mu^{2l^*} < 1$ and take any $c^\xi>0$. Let $c_3=\sum_{j=0}^{n-1} \left(2 \bar f^2 \mu^{2l^*} \right)^j$. Then, taking $\gamma=\left(c_2 +c^\xi/c_3\right)/(1- 2 \bar f^2 \mu^{2l^*})$ leads to  
    $V_{k+n} - V_k \leq -c^\dagger \|\msf{e}_k^\dagger\|^2 -c^\xi \|\breve{\msf{e}}_k^\xi\|^2.$
From this, the asymptotic stability is obtained with a Lyapunov function $\mathcal V_k = \sum_{j=0}^{n-1} V_{nk+j}$.
This completes the proof.  \hfill $\square$
\end{pf}
\vspace{-0.5cm}

\section{Numerical Example}\label{sec:numerical_experiments}\vspace{-0.2cm}

Consider a collectively observable sensor network consisting of one hundred estimators. The system matrix of target system is given by $F = \diag\{F_1, F_2\}$ where $F_1 = \begin{bmatrix}
0.4 & 0.9 \\ -0.9 & 0.4
\end{bmatrix}$ and
$F_2 = \begin{bmatrix}
0.5 & 0.8 \\ -0.8 & 0.5
\end{bmatrix}$. 
The output matrices associated with the sensors, $H_i$, are chosen randomly where each element has $-1, 0,$ or $1$. It is assumed that  $Q = 0.05 I_2$ and $\bar R = 0.05 I_{100}$. The network topology is depicted in Fig. \ref{fig:net4} and all the edge weights are $1$. The maximum eigenvalue of the Laplacian matrix associated to the network is $14.26$. $\alpha_{\lambda,i}$ and $\alpha_{\upsilon,i}$ are chosen as $0.009$, and $\epsilon_i = 1, \forall i \in \mathcal N$. The initial estimate $\hat x_{i,0}$ is randomly selected, and we choose the initial covariance of each estimator as $I_4$. 

Fig. \ref{fig:exp1-1} presents the estimation errors of the filters with $l^*=1$. As shown in the figure, the estimation error for each filter is bounded in the steady-state with small error. The error between the local covariance and the steady-state covariance of CKF converges to zero. Fig. \ref{fig:exp1-2} shows the effect of $l^*$ on the estimation performance. This result has been obtained through $100$ repeated experiments, and in all cases ($l^* = 1$ to $l^* = 7$) the mean squared error (MSE) $\frac{1}{N} \sum_{i=1}^N \|x_k - \hat x_{i,k}\|^2$ remains close to zero in the steady-state. The convergence performance improves as $l^*$ increases. In all cases, MSE of covariance coverges to zero. 
\begin{figure}[t!]
\centering
\includegraphics[scale=0.25]{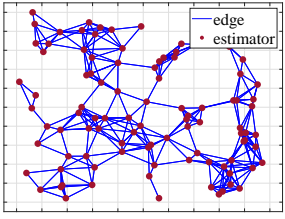} 
\caption{Network topology consisting of one hundred nodes.}
\label{fig:net4}
\end{figure}

\begin{figure}[t!]
\begin{subfigure}{.5\textwidth}
\centering
\!\!\!\includegraphics[scale=0.35]{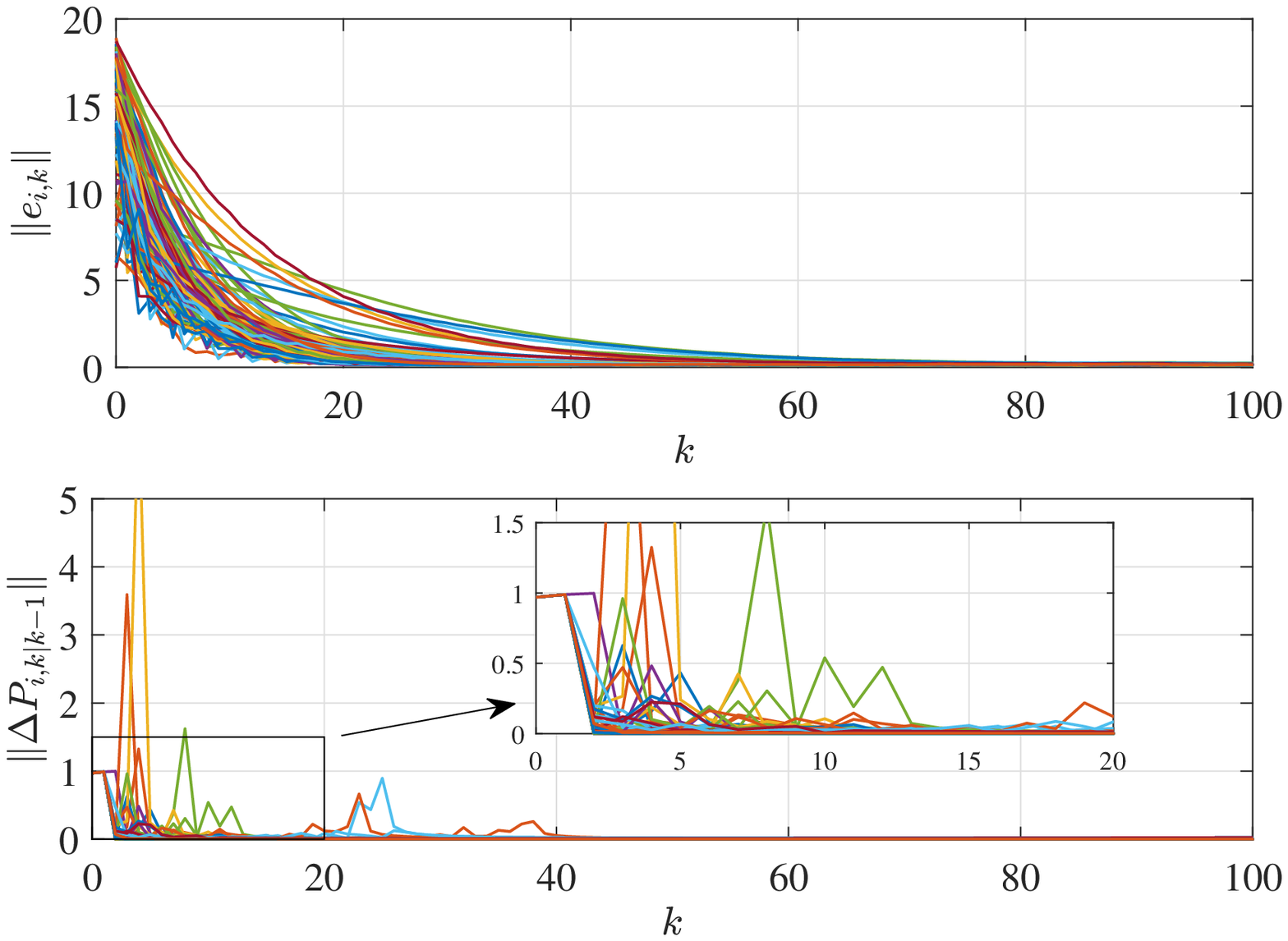} 
\end{subfigure}
\vspace{-0.3cm}
\caption{Estimation results of DA-DKF ($l^* = 1$): norm of estimation error (top) and norm of covariance error (bottom).} \label{fig:exp1-1}
\begin{subfigure}{.5\textwidth}
\centering
\!\!\!\includegraphics[scale=0.35]{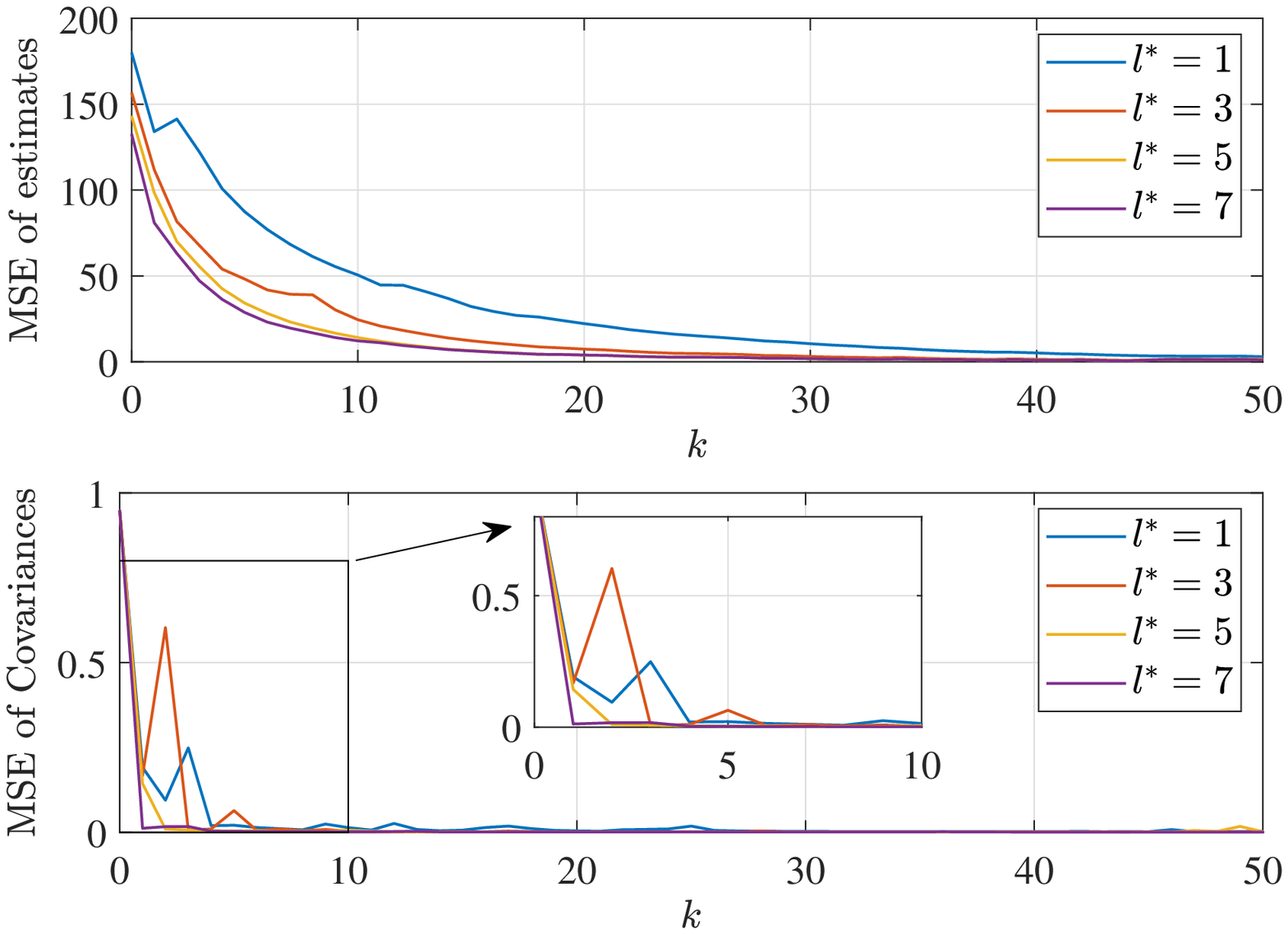} 
\end{subfigure}
\vspace{-0.3cm}
\caption{Estimation performance of DA-DKF when $l^*$ varies:  estimates (top) and  covariances (bottom).} \label{fig:exp1-2}
\end{figure}

\section{Conclusion}\label{sec:conclusion}\vspace{-0.2cm}
In this paper, we have formulated the DKF problem as a consensus optimization problem. It is expected that this new perspective enables us to develop novel DKF algorithms by employing various efficient distributed optimization techniques.
As an instance, we have proposed DA-DKF, adopting the dual-ascent method. The unbiased property and the convergence of local covariances to the steady-state covariance of CKF are shown, and this implies that DA-DKF recovers the performance of CKF asymptotically.

Formulating a problem as an optimization problem enables us to deal with constraints very efficiently. This implies that DKF with constraints can be effectively handled under the proposed formulation, which is one of our future research topics. Another important research direction is to consider more general network topology such as directed and time varying graph, etc. In addition, relaxing the assumptions such as the known number of estimators or extending to collective detectability will improve the applicability of the proposed approach.
\vspace{-0.2cm}
\section*{Appendices}
\appendix{\section{Proof of Lemma \ref{lem:cov_rule}}\label{appdx:cov_rule}}
For convenience, we proceed using $\Theta_{i,k,l} := \invVEC{\theta_{i,k,l}}$ and $\Upsilon_{i,k,l} := \invVEC{\upsilon_{i,k,l}}$ rather than $\theta_{i,k,l}$ and $\upsilon_{i,k,l}$. 
Let $\Theta_{k,l} = [\Theta_{1,k,l}; \cdots; \Theta_{N,k,l}] \in R^{Nn \times n}$ and $\Upsilon_{k,l} = [\Upsilon_{1,k,l}; \cdots; \Upsilon_{N,k,l}] \in R^{Nn \times n}$. For the case $k \geq 1$ and $l \geq 1$, it follows from \eqref{eq:cov_rule} that 
\begin{equation}\label{eq:theta_iteration}
\begin{split}
    \Theta_{k,l+1} &= N \Omega^\delta - {\mathbb L} \Upsilon_{k,l+1}
    = ({\mathbb I}_N - {\mathbb L} \bar A_\upsilon {\mathbb L}) \Theta_{k,l}
\end{split}
\end{equation}
where $\Omega^\delta := [\Omega^\delta_1;\cdots;\Omega^\delta_N]$ with $\Omega^\delta_i := \invVEC{\omega^\delta_i} = H_i^\top R^{-1}_i H_i$. 

Recall that $\Theta^\dagger = H^\top \bar R^{-1} H$ from Lemma \ref{lem:opt_pt_cov}, and let $e_{i,k,l}^\Theta := \Theta^\dagger - \Theta_{i,k,l}$ and $e_{k,l}^\Theta :=[e_{1,k,l}^\Theta; \cdots; e_{N,k,l}^\Theta]$. Then, from \eqref{eq:theta_iteration}, we have $e^\Theta_{k,l+1} = ({\mathbb I}_N - {\mathbb L} \bar A_\upsilon {\mathbb L} ) e^\Theta_{k,l}$.

Applying the coordinate transformation given by $\breve e^\Theta_{k,l} = {\mathbb U}^\top e^\Theta_{k,l}$, we have 
$\breve e_{k,l+1}^\Theta = ({\mathbb I}_N - {\mathbb \Lambda} {\mathbb U}^\top \bar A_\upsilon {\mathbb U} {\mathbb \Lambda}) \breve e_{k,l}^\Theta = \diag\{I_n, {\mathbb I}_{N-1} - \tilde {\mathbb \Lambda} {\mathbb W}^\top \bar A_\upsilon {\mathbb W} \tilde {\mathbb \Lambda}\} \breve e_{k,l}^\Theta$.
If we choose $\alpha_{\upsilon,i}$ satisfying \eqref{eq:alpha_ups}, it holds that $\tilde {\mathbb \Lambda} {\mathbb W}^\top \bar A_\upsilon {\mathbb W} \tilde {\mathbb \Lambda} < 2{\mathbb I}_{N-1}$. Since $\tilde {\mathbb \Lambda} {\mathbb W}^\top \bar A_\upsilon {\mathbb W} \tilde {\mathbb \Lambda}$ is symmetric positive definite, we have ${\mathbb I}_{N-1} - \tilde {\mathbb \Lambda} {\mathbb W}^\top \bar A_\upsilon {\mathbb W} \tilde {\mathbb \Lambda}$ is Schur stable.  

Meanwhile, the elements of the first $n$ rows of $\breve e_{k,l}^\Theta$ remain the same for any $l \geq 1$, namely, ${\mathbb 1}_N^\top e^\Theta_{k,l} = {\mathbb 1}_N^\top e^\Theta_{1,1}= {\mathbb 1}_N^\top ({\mathbb 1}_N \Theta^\dagger - N\Omega^\delta - {\mathbb L} \Upsilon_{1,1} ).$ 
Since $\Theta^\dagger = \sum_{i=1}^N H_i^\top R_i^{-1} H_i$ (Lemma \ref{lem:opt_pt_cov}) and ${\mathbb 1}_N^\top \Omega^\delta = \sum_{i=1}^N H_i^\top R_i^{-1} H_i$, it follows that the first $n$ rows of $\breve e_{k,l}^\Theta$ is zero, i.e., ${\mathbb 1}_N^\top e^\Theta_{k,l} = 0$. 
Thus, we have $\breve e_{k,l^*}^\Theta  = ({\mathbb I}_N - {\mathbb \Lambda} {\mathbb W}^\top \bar A_\upsilon {\mathbb W} {{\mathbb \Lambda}})^{l^*} \breve e_{k,0}^\Theta
=({\mathbb I}_N - {\mathbb \Lambda} {\mathbb U}^\top \bar A_\upsilon {\mathbb U} {{\mathbb \Lambda}})^{kl^*} \breve e_{1,0}^\Theta= [0_{n\times n}; ({\mathbb I}_{N-1} - \tilde{\mathbb \Lambda} {\mathbb W}^\top \bar A_\upsilon {\mathbb W} \tilde{\mathbb \Lambda} )^{kl^*} {\mathbb W}^\top e_{1,0}^\Theta]$, 
from which we conclude that $e^\Theta_{k,l^*}$ converges to zero as $k$ goes to infinity for any $l^* \geq 1$.

We prove the second part, the existence of $\bar k$. Suppose $k > 1$. From \eqref{eq:theta_iteration}, one has $\Theta_{k,l+1} = ({\mathbb I}_N - {\mathbb L} \bar A_\upsilon {\mathbb L})^{kl^*-1} \Theta_{1,1}$. From \eqref{eq:cov_rule}, $\Upsilon_{1,1}$ and $\Theta_{1,1}$ are computed with $k=1$ and $l=0$ as $\Upsilon_{1,1} = \bar A_\upsilon {\mathbb L} \Theta_{1,0}$ and $\Theta_{1,1} = N \Omega^\delta - {\mathbb L} \Upsilon_{1,1}$ which results in
\begin{equation*}
\begin{split}
    \Theta_{1,1} = N \Omega^\delta - {\mathbb L} \bar A_\upsilon {\mathbb L} \Theta_{1,0}.
\end{split}
\end{equation*}
Since $\Omega^\delta = \Theta_{1,0}$ by Algorithm \ref{algo:DA_DKF}, we have $\Theta_{1,1} = (N \mathbb I - {\mathbb L} \bar A_\upsilon {\mathbb L}) \Theta_{1,0}$. Recalling that $-\mathbb I < (\mathbb I - {\mathbb L} \bar A_\upsilon {\mathbb L}) < \mathbb I$, we have $-N \mathbb I < N \mathbb I - {\mathbb L} \bar A_\upsilon {\mathbb L} < N \mathbb I$. In addition, from the fact that $\Theta_{i,1,0} < \Theta^\dagger, \forall i \in \mathcal N$, we obtain $\Theta_{i,1,1} < N \Theta^\dagger$. Since $\mathbb I - {\mathbb L} \bar A_\upsilon {\mathbb L}$ is Schur stable, there exists $0 < \rho < 1$ such that 
\begin{equation} \label{eq:rho_kl_star}
   -N \rho^{kl^*-1} \Theta^\dagger \leq e_{i,k,l^*}^\Theta \leq N \rho^{kl^*-1} \Theta^\dagger.
\end{equation}
Hence, for any $0 <\kappa < 1$, there exists $\bar k$ such that 
    $- \kappa \Theta^\dagger \leq e^\Theta_{i,k,l^*} \leq \kappa \Theta^\dagger, \  \forall k \geq \bar k$,
and by recalling that $e^\Theta_{i,k,l^*} = \Theta^\dagger - \Theta_{i,k,l^*}$, we have \eqref{eq:Theta_ulbound}. This completes the proof. 

\section{Proof of Lemma \ref{lem:cov_bound}}\label{appdx:cov_bound}

Let $\kappa$ be given such that $0<\kappa<1$. By Lemma \ref{lem:cov_rule}, there exists $\bar k$ such that the relation \eqref{eq:Theta_ulbound} holds true.

1) Existence of $\overline P$:  The proof is done by exploiting the monotonicity of the algebraic Riccati equation \cite{Bitmead+1985SCL}. 

We take a symmetric positive definite matrix $P_{\bar k}^\sharp$ such that $P_{i, k} \leq P_{\bar k}^\sharp$ for all $i\in \mathcal N$ and for all $k\le\bar k$. Then, 
\begin{equation}\label{eq:P_is_bounded_by_P_sharp}
    P_{i, k+1| k}\leq P_{\bar k+1| \bar k}^\sharp, \quad \forall i \in \mathcal N, \forall k\le \bar k
\end{equation}
where $P_{i, k+1| k} \!=\! F P_{i, k} F^\top \!+ Q$ and $P_{\bar k+1| \bar k}^\sharp \!=\! F P_{\bar k}^\sharp F^\top + Q$.

Meanwhile, let $\overline{P}$ be the unique positive definite solution to the algebraic Riccati equation given by
\begin{equation}\label{eq:overline_P}
    \overline{P} = F \left(\overline{P}^{-1} + (1-\kappa) \Theta^\dagger \right)^{-1} F^\top + \bar qI
\end{equation}
where $\bar q$ is a positive scalar such that $P_{\bar k+1| \bar k}^\sharp \leq \bar q I$. The existence of $\overline P$ is guaranteed by the controllability of $(F, \sqrt{\bar q} I)$ and the observability of $(F, \sqrt{1-\kappa} H)$ \cite{Bitmead+1985SCL}. 
From the manner in which $\bar q$ is chosen, it follows from \eqref{eq:overline_P} that $P_{\bar k+1| \bar k}^\sharp \leq \bar qI \leq \overline{P}$.
Then, using the relations \eqref{eq:Theta_ulbound}, \eqref{eq:P_is_bounded_by_P_sharp}, \eqref{eq:overline_P}, $P_{\bar k+1| \bar k}^\sharp \leq  \overline{P}$, and $Q \leq \bar qI$, we derive  
\begin{equation*}
\begin{split}
P_{i,\bar k+2 | \bar k+1} &= F ( P_{i,\bar k+1 | \bar k}^{-1} + \Theta_{i,\bar k+1,l^*} )^{-1} F^\top + Q\\
    & \leq F \left( P_{\bar k+1 | \bar k}^{\sharp ~ -1} + \Theta_{i,\bar k+1,l^*} \right)^{-1} F^\top + Q\\
    & \leq F \left(\overline{P}^{-1} + (1-\kappa) \Theta^\dagger \right)^{-1} F^\top + \bar qI \\
    &= \overline{P}.
\end{split}
\end{equation*}
Suppose that $P_{i,\bar k+j|\bar k + j -1} \leq \overline P$ holds for a particular $j \geq 0$. Then, it follows that 
$P_{i,\bar k+j+1 | \bar k+j} \leq \overline{P}$,
which proves that $P_{i,k|k-1}\le \overline P$, $\forall k \ge 1$ and $i\in \mathcal N$. The bound for $P_{i,k+1}$ is the same since for any $k \ge 1$ and $i\in \mathcal N$ it holds that
     $P_{i,k+1} = (P_{i,k+1|k}^{-1} + \Theta_{i,k+1,l^*})^{-1} \leq P_{i,k+1|k} \leq \overline{P}$.

2) Existence of $\underline P$: From the prediction rule of covariance given by $P_{i,k|k-1} = F P_{i,k-1} F^\top + Q$, it follows that 
\begin{equation} \label{eq:prior_cov_lower_bound}
    Q \leq P_{i,k|k-1}.    
\end{equation}
Moreover, since the prior covariance $P_{i,k}$ is obtained as $ P_{i,k} = (P_{i,k|k-1}^{-1} + \Theta_{i,k,l^*})^{-1}$,
we have from \eqref{eq:Theta_ulbound} and \eqref{eq:prior_cov_lower_bound}  that
$P_{i,k} \geq (Q^{-1} + (1+\kappa)\Theta^\dagger)^{-1}, \quad \forall i \in \mathcal N, \forall k \ge \bar k$,
which proves the existence of $\underline P$. 

\section{Proof of Lemma \ref{lem:est_rule}}\label{appdx:est_rule}
We first derive the dynamics of $e_{k,l}^\lambda := \lambda_k^* - \lambda_{k,l}$. 
From \eqref{eq:est_rule}, we have
\begin{equation*}
\begin{split}
e^\lambda_{k,l+1} &= \lambda^*_k - \lambda_{k,l} - \bar A_\lambda K_k^{\mathsf{dual}} {\mathbb L} \xi_{k,l}\\
&= e^\lambda_{k,l} - \bar A_\lambda K_k^{\mathsf{dual}} {\mathbb L} K_k^{\mathsf{cons}} (\diagg{\augH}^\top \diagg{\augS}_k^{-1} \augz_k - {\mathbb L}\lambda_{k,l})
\end{split}
\end{equation*}
where $\xi_{k,l} := [\xi_{k,1}; \cdots; \xi_{k,N}]$ and $\lambda_{k,l} := [\lambda_{k,1}; \cdots; \lambda_{k,N}]$.
From the dual feasibility equation in  \eqref{eq:SPE}, we have 
\begin{equation}\label{eq:xi_*}
\xi_k^* = K_k^{\mathsf{cons}} \left( \diagg{\augH}^\top \diagg{\augS}_k^{-1} \augz_k - {\mathbb L} \lambda^*_k \right).  
\end{equation}

Since ${\mathbb L} \xi_k^* = 0$, adding $\bar A_\lambda K_k^{\mathsf{dual}} {\mathbb L} \xi_k^*$ to $e^\lambda_{k,l+1}$ yields 
\begin{equation}\label{eq:e_lambda_l_recursive}
e^\lambda_{k,l+1} = \left({\mathbb I}_N - \bar A_\lambda K_k^{\mathsf{dual}}  {\mathbb L} K_k^{\mathsf{cons}}{\mathbb L} \right) e^\lambda_{k,l}.
\end{equation} 
    
Meanwhile, Theorem \ref{thm:cov_convergence} guarantees that for any $\epsilon_{\mathsf P} > 0$ such that $P^*-\epsilon_{\mathsf P}I_n>0$ and $\epsilon_{\mathsf P}< \epsilon_i/(2N), \forall i\in \mathcal N$, there exists $k^*$ such that $P^*-\epsilon_{\mathsf P}I_n < P_{i,k|k-1}< P^* +\epsilon_{\mathsf P} I_n$, for all $k\ge k^*$ and $i \in \mathcal N$. This means that $P_{i,k|k-1}$ converges to $P^*$ as $k$ goes infinity. Hence, by using $P^*-\epsilon_{\mathsf P}I_n < P_{i,k|k-1}< P^* +\epsilon_{\mathsf P} I_n$ and $2N \epsilon_{\mathsf P}< \epsilon_i, \forall i\in \mathcal N$, we have for all $k\ge k^*$ that
\begin{align*}
        \|K_k^{\mathsf{dual}} \| &\le \max_{i \in \mathcal N} \frac{1}{\|N P_{i,k|k-1}\| +\epsilon_i} \le \frac{1}{\|N P^* + N \epsilon_{\mathsf P}I_n \|}.
    \end{align*}
Similarly, from the facts that $(H_i^\top R_i^{-1} H_i + \frac{1}{N} P_{i,k|k-1}^{-1})^{-1} \leq N P_{i,k|k-1}$ and $P_{i,k|k-1} < P^* + \epsilon_{\mathsf P} I_n$, one has for any $k\ge k^*$ that $\|K_k^{\mathsf {cons}}\| \leq  \|N P^* + N \epsilon_{\mathsf P} I_n\|$, $\forall k\ge k^*$.
Applying the bounds for $\|K_k^{\mathsf{cons}}\|$ and $\|K_k^{\mathsf{dual}}\|$, and recalling  $\|{\mathbb L}\| \leq \bar \sigma$, we have $\|\bar A_\lambda K_k^{\mathsf{dual}}  {\mathbb L} K_k^{\mathsf{cons}}{\mathbb L}\| < 2$ if $\alpha_{\lambda,i}$ satisfies \eqref{eq:alpha_upsilon_lambda}.  
Since the matrices $\bar A_\lambda, K_k^{\mathsf{cons}}, K_k^{\mathsf{dual}}$, and $\mathbb L$ are symmetric and positive semidefinite, it follows that all the eigenvalues of $\bar A_\lambda K_k^{\mathsf{dual}}  {\mathbb L} K_k^{\mathsf{cons}}{\mathbb L}$ are nonnegative real numbers. This fact and the property $\|\bar A_\lambda K_k^{\mathsf{dual}}  {\mathbb L} K_k^{\mathsf{cons}}{\mathbb L}\| < 2$ result in that all the eigenvalues of ${\mathbb I}_N - \bar A_\lambda K_k^{\mathsf{dual}}  {\mathbb L} K_k^{\mathsf{cons}}{\mathbb L}$, except the ones at 1, are located inside the unit circle, provided that $\alpha_{\lambda,i}$ is chosen as \eqref{eq:alpha_upsilon_lambda}.

We now compute the limit of $e^{\lambda}_{k,l}$ as $l \rightarrow \infty$. 
From the fact that $K_k^{\mathsf{dual}} {\mathbb L} K_k^{\mathsf{cons}}{\mathbb L}$ is composed of symmetric positive semidefinite matrices and that $\mathbb L$ has $n$ eigenvalues at zero, it follows that there exists a nonsingular matrix $\hat U_k$ of the form $\hat U_k = [\mathsf u \otimes I_n ~\hat W]$ such that $\bar A_\lambda K_k^{\mathsf{dual}} {\mathbb L} K_k^{\mathsf{cons}} {\mathbb L} \hat U_k = \hat U_k \diag\{0_{n\times n}, \hat \Lambda_k\}$ where $\hat \Lambda_k$ is a diagonal matrix. Then, in the new coordinates $\hat e^\lambda_{k,l} = \hat U_k^{-1} e^\lambda_{k,l}$, we have 
$\hat e^\lambda_{k,l+1} =  \diag\{I_n, {\mathbb I}_{N-1} - \hat \Lambda_k\}  \hat e^\lambda_{k,l}.$

From the discussion on the eigenvalues, it follows that ${\mathbb I}_{N-1} - \hat \Lambda_k$ becomes Schur for $\alpha_{\lambda,i}$ satisfying \eqref{eq:alpha_upsilon_lambda}, which results in that 
    $\lim_{l \rightarrow \infty} \hat e^\lambda_{k,l} = \diag\{I_n, 0_{(N-1)n \times n}\} \hat e^\lambda_{k,0}$,
equivalently,
\begin{equation}\label{eq:lim_e_lambda}
    \lim_{l \rightarrow \infty} e^\lambda_{k,l} = \frac{1}{\sqrt{N}} {\mathbb 1}_N \hat e^\lambda_{k,0}.
\end{equation}

To prove the convergence of $\xi_{k,l}$ to $\xi_k^*$, define $e_{k,l}^\xi=\xi_{k}^*-\xi_{k,l}$. From \eqref{eq:est_rule_primal} and \eqref{eq:xi_*}, one has
\begin{equation}\label{eq:e_xi_to_e_lambda}
    e_{k,l}^\xi = - K_k^{\mathsf{cons}}{\mathbb L}e_{k,l}^\lambda,
\end{equation}
and by applying \eqref{eq:lim_e_lambda}  we obtain that $\lim_{l \rightarrow \infty} e^\xi_{k,l} = 0.$

We now prove the existence of $\mu$ such that $\|\tilde{\Xi}_k\| < \mu < 1$. To do this, we derive the dynamics of $e_{k,l}^\xi$. Substituting \eqref{eq:e_lambda_l_recursive} to \eqref{eq:e_xi_to_e_lambda} yields
\begin{equation}\label{eq:e_xi_dyn}
\begin{split}
    e^\xi_{k,l+1} &= \left({\mathbb I}_N  - K_k^{\mathsf{cons}}{\mathbb L} \bar A_\lambda K_k^{\mathsf{dual}} {\mathbb L} \right) e^\xi_{k,l}=: \Xi_k e^\xi_{k,l}.
\end{split}
\end{equation}
Let $\breve{e}_{k,l }^\xi = {\mathbb U}^\top e_{k,l }^\xi$. Then, the dynamics of  $\breve{e}_{k,l }^\xi$ reads as
\begin{equation*}
\begin{split}
    \breve{e}_{k,l+1}^\xi &= \left({\mathbb I}_N - {\mathbb U}^\top K_k^{\mathsf{cons}} {\mathbb U} {\mathbb \Lambda} {\mathbb U}^\top  \bar A_\lambda K_k^{\mathsf{dual}} {\mathbb U} {\mathbb \Lambda} \right) \breve{e}^\xi_{k,l}=: \breve{\Xi}_k \breve{e}^\xi_{k,l}.
\end{split}
\end{equation*}
Recalling that ${\mathbb U} = [{\mathsf u}\otimes I_n ~ {\mathbb W}]$, one can write $\breve{\Xi}_k$ as
\begin{equation*}
\begin{split}
    \breve{\Xi}_k =& \begin{bmatrix}
    I_n & \check{\Xi}_k\\
    0 & \tilde{\Xi}_k
    \end{bmatrix}
\end{split}
\end{equation*}
where $\check{\Xi}_k = - ({\mathsf u}^\top \otimes I_n) K_k^{\mathsf{cons}} {\mathbb W} \tilde{\mathbb \Lambda} {\mathbb W}^\top \bar A_\lambda K_k^{\mathsf{dual}} {\mathbb W} \tilde{\mathbb \Lambda}$ and 
$\tilde{\Xi}_k = {\mathbb I}_{N-1} - {\mathbb W}^\top K_k^{\mathsf{cons}} {\mathbb W} \tilde{\mathbb \Lambda} {\mathbb W}^\top \bar A_\lambda K_k^{\mathsf{dual}} {\mathbb W} \tilde{\mathbb \Lambda}$. 

Since $\|{\mathbb W}\| = 1$, following the same reasoning as \eqref{eq:e_lambda_l_recursive}, one can show that $\|\tilde{\Xi}_k\| < 1$, and this results in the existence of a positive scalar $\mu$ such that $\|\tilde{\Xi}_k\| < \mu < 1$, $\forall k \geq k^*$. This completes the proof. 
\section{Proof of Lemma \ref{lem:error_dyn}}\label{appdx:error_dyn}
Recalling that $\mathcal{H}_k = {\mathbb 1}_N^\top (K_k^{\mathsf{cons}})^{-1} {\mathbb 1}_N $ from \eqref{eq:define_z_H_S} and $\xi_k^\dagger = \mathcal{H}_k^{-1} {\mathbb 1}_N^\top \diagg{\augH}^\top \diagg{\augS}_k^{-1} \augz_k$ from \eqref{eq:xi_k_dagger}, we have
\begin{align}\label{eq:e_dagger}
{\mathsf e}_{k+1}^\dagger
&= \mathcal{H}_{k+1}^{-1} \mathcal{H}_{k+1} \mathbb E\{x_{k+1}\} - \mathcal  H_{k+1}^{-1} {\mathbb 1}_N^\top \mathbb E\{ \diagg{\augH}^\top \diagg{\augS}_{k+1}^{-1} \augz_{k+1}\} \nonumber \\
&= \mathcal{H}_{k+1}^{-1} {\mathbb 1}_N^\top \big(  (K_{k+1}^{\mathsf{cons}})^{-1} {\mathbb 1}_N \mathbb E\{x_{k+1}\} \nonumber \\
& \quad - \mathbb E\{ \diagg{\augH}^\top \diagg{\augS}_{k+1}^{-1} \augz_{k+1}\} \big).
\end{align}
Since $K_{k+1}^{\mathsf{cons}} = (\diagg{H}^\top \diagg{R}^{-1} \diagg{H} + \frac{1}{N} \bar P_{k+1|k}^{-1})^{-1}$ , 
the first term in parentheses in \eqref{eq:e_dagger} can be rearranged as follows.
\begin{equation}\label{eq:HSHx}
\begin{split}
    & (K_{k+1}^{\mathsf{cons}})^{-1} {\mathbb 1}_N \mathbb E\{x_{k+1}\} \\
    & = \diagg{H}^\top \diagg{R}^{-1} \diagg{H} {\mathbb 1}_N\mathbb E\{x_{k+1}\} + \frac{1}{N} \bar P_{k+1|k}^{-1} {\mathbb 1}_N F \mathbb E\{x_{k}\}.
\end{split}
\end{equation}
For the second term in parentheses in \eqref{eq:e_dagger}, we have
\begin{equation}\label{eq:HSz}
\begin{split}
    &\mathbb E\{ \diagg{\augH}^\top \diagg{\augS}_{k+1}^{-1} \augz_{k+1} \}\\ 
    & = \diagg{H}^\top \diagg{R}^{-1} \diagg{H} {\mathbb 1}_N  \mathbb E\{x_{k+1}\}  + \frac{1}{N} \bar P_{k+1|k}^{-1} \mathbb E\{\hat x_{k+1|k}\}.
\end{split}
\end{equation}
Subtracting \eqref{eq:HSz} from \eqref{eq:HSHx} and applying the identities $\hat x_{k+1|k} = \diagg{F} \xi_{k,l^*}$ and ${\mathbb 1}_N F \mathbb E\{x_k\} - \diagg{F} \mathbb E \{\xi_{k,l^*}\} \pm  {\mathbb 1}_N F\xi_k^\dagger = {\mathbb 1}_N F \msf{e}_k^\dagger + \diagg{F} \msf{e}_{k,l^*}^\xi$, we have $(K_{k+1}^{\mathsf{cons}})^{-1} {\mathbb 1}_N \mathbb E\{x_{k+1}\}\!-\!\mathbb E\{\diagg{\augH}^\top \diagg{\augS}_{k+1}^{-1} \augz_{k+1}\} = \frac{1}{N} \bar P_{k+1|k}^{-1} \left({\mathbb 1}_N F\msf{e}_k^\dagger\!+\!\diagg{F}  \msf{e}^\xi_{k,l^*} \right)$. 
Then, \eqref{eq:e_dagger} becomes
\begin{align}\label{eq:e_k+1_dagger_to_e_k_dagger}
    \msf{e}^\dagger_{k+1} &= E_{k+1}^\dagger  \msf{e}_k^\dagger + \mathcal{H}_{k+1}^{-1} {\mathbb 1}_N^\top \frac{1}{N} \bar P_{k+1|k}^{-1} \diagg{F} \msf{e}^\xi_{k,l^*}.
\end{align}
We now derive the dynamics of $\msf{e}_{k+1,l^*}^\xi$. It holds from \eqref{eq:e_xi_dyn} that $\msf{e}_{k+1,l^*}^\xi = (\Xi_{k+1})^{l^*} \msf{e}_{k+1,0}^\xi$. Recalling that $\xi_k^* = {\mathbb 1}_N  \mathcal{H}_k^{-1} {\mathbb 1}_N^\top \diagg{\augH}^\top \diagg{\augS}_k^{-1} \augz_k$, one can obtain
\begin{equation}\label{eq:e_k+1_xi}
\begin{split}
   \!\!\!  \msf{e}_{k+1,0}^\xi\! &= \mathbb E\{\xi_{k+1}^* - \xi_{k+1,0}\}\\
    &= {\mathbb 1}_N  \mathcal{H}_{k+1}^{-1} {\mathbb 1}_N^\top \mathbb E\{\diagg{\augH}^\top \diagg{\augS}_{k+1}^{-1} \augz_{k+1}\}\! - \mathbb E\{\xi_{k+1,0}\}.\!
\end{split}
\end{equation}
From the fact that $\hat x_{i,k+1|k} = \xi_{i,k+1,0} = F \xi_{i,k,l^*}$ and the relation \eqref{eq:HSz}, we have $\mathbb E\{\diagg{\augH}^\top \diagg{\augS}_{k+1}^{-1} \augz_{k+1}\} = \diagg{H}^\top \diagg{R}^{-1} \diagg{H}{\mathbb 1}_N F  \mathbb E\{x_{k}\} + \frac{1}{N} \bar P_{k+1|k}^{-1}   \diagg{F} \mathbb E\{\xi_{k,l^*}\}$,
and substituting this identity to \eqref{eq:e_k+1_xi} gives
\begin{align*}
     \msf{e}_{k+1,0}^\xi =& {\mathbb F}_{ k}  \mathbb E\{x_k\}- {\mathbb G}_{k} \bar F\mathbb E\{\xi_{k,l^*}\},
\end{align*}
where ${\mathbb F}_{k}={\mathbb 1}_N  \mathcal{H}_{k+1}^{-1} {\mathbb 1}_N^\top \diagg{H}^\top \diagg{R}^{-1} \diagg{H} {\mathbb 1}_N F$ and ${\mathbb G}_{k}={\mathbb I}_N - {\mathbb 1}_N  \mathcal{H}_{k+1}^{-1} {\mathbb 1}_N^\top \frac{1}{N}\bar P_{k+1|k}^{-1}$. Adding and subtracting $
    {\mathbb F}_{k}\mathbb E\{\xi_{k}^\dagger\}$, and applying the identity ${\mathbb G}_k{\mathbb 1}_N   = {\mathbb 1}_N  \mathcal{H}^{-1}_{k+1} {\mathbb 1}_N^\top \diagg{H}^\top \diagg{R}^{-1} \diagg{H} {\mathbb 1}_N$, we have 
$   \msf{e}_{k+1,0}^\xi ={\mathbb F}_k \msf{e}_k^\dagger + {\mathbb G}_k   \diagg{F} \msf{e}^\xi_{k,l^*}$.From $\msf{e}_{k+1,l^*}^\xi = (\Xi_{k+1})^{l^*} \msf{e}_{k+1,0}^\xi$, it holds that 
$    \msf{e}_{k+1,l^*}^\xi = (\Xi_{k+1})^{l^*} {\mathbb F}_k \msf{e}_k^\dagger  + (\Xi_{k+1})^{l^*} {\mathbb G}_k \diagg{F} \msf{e}^\xi_{k,l^*}$,
and applying the fact that $\Xi_{k+1} {\mathbb 1}_N  = {\mathbb 1}_N $ yields 
\begin{equation}\label{eq:e_k+1_xi_to_e_k_xi_reduced_1}
\begin{split}
 \!\!\!\! \!\!\!\!  & \msf{e}_{k+1,l^*}^\xi = {\mathbb 1}_N  \mathcal{H}_{k+1}^{-1} \underline{{\mathbb 1}_N^\top \diagg{H}^\top \diagg{R}^{-1} \diagg{H} {\mathbb 1}_N F \msf{e}_k^\dagger} \\
  \!\!  &   \qquad + \left( (\Xi_{k+1})^{l^*} \!-\! {\mathbb 1}_N  \mathcal{H}_{k+1}^{-1} {\mathbb 1}_N^\top \frac{1}{N}\bar P_{k+1|k}^{-1} \right)\!\diagg{F}\msf{e}^\xi_{k,l^*}.\!\!\!
\end{split}
\end{equation}
Meanwhile, the dynamics \eqref{eq:e_k+1_xi_to_e_k_xi_reduced_1} has a constraint 
$-{\mathbb L} \msf{e}_{k+1,l^*}^\lambda = (K_{k+1}^{\mathsf{cons}})^{-1} \msf{e}_{k+1,l^*}^\xi$,
which comes from the dual feasibility equation in  \eqref{eq:SPE} and the update rule \eqref{eq:est_rule_primal} (or from \eqref{eq:e_xi_to_e_lambda}).  
Multiplying the left side of this constraint by ${\mathbb 1}_N^\top$ gives ${\mathbb 1}_N^\top  (K_{k+1}^{\mathsf{cons}})^{-1} \msf{e}_{k+1,l^*}^\xi = 0$.
Then, substituting \eqref{eq:e_k+1_xi_to_e_k_xi_reduced_1} into the above equation and recalling that $\mathcal{H}_{k} = {\mathbb 1}_N^\top (K_{k}^{\mathsf{cons}})^{-1}  {\mathbb 1}_N $, we obtain ${\mathbb 1}_N^\top \diagg{H}^\top \diagg{R}^{-1} \diagg{H} {\mathbb 1}_N F \msf{e}_k^\dagger=- {\mathbb 1}_N^\top (K_{k+1}^{\mathsf{cons}})^{-1} (\Xi_{k+1})^{l^*} \diagg{F} \msf{e}_{k,l^*}^\xi + {\mathbb 1}_N^\top \frac{1}{N} \bar P_{k+1|k}^{-1} \diagg{F}\msf{e}_{k,l^*}^\xi$. Substituting this into the underlined part of \eqref{eq:e_k+1_xi_to_e_k_xi_reduced_1} yields
\begin{align*}
    \msf{e}_{k+1,l^*}^\xi &= \left( {\mathbb I}_N - {\mathbb 1}_N \mathcal{H}_{k+1}^{-1} {\mathbb 1}_N^\top  (K_{k+1}^{\mathsf{cons}})^{-1} \right) (\Xi_{k+1})^{l^*} \diagg{F} \msf{e}_{k,l^*}^\xi \\
    &=: E^\xi_{k+1} (\Xi_{k+1})^{l^*} \diagg{F} \msf{e}_{k,l^*}^\xi.
\end{align*}
Consider the change of variables $\breve{\msf{e}}^\xi_{k,l^*} = {\mathbb U}^\top \msf{e}^\xi_{k,l^*}$. Then, the dynamics in the new coordinates becomes
\begin{equation}\label{eq:bar_e_k+1_xi_to_e_k_xi}
\begin{split}
    \breve{\msf{e}}_{k+1,l^*}^\xi &= \breve{E}_{k+1}^{\xi} (\breve{\Xi}_{k+1})^{l^*} \diagg{F} \breve{\msf{e}}_{k,l^*}^\xi 
\end{split}
\end{equation}
where $\breve{E}_{k+1}^{\xi} = {\mathbb U}^\top E_{k+1}^{\xi} {\mathbb U}$ and $\breve{\Xi}_{k} = {\mathbb U}^\top \Xi_{k} {\mathbb U}$. With \eqref{eq:e_k+1_dagger_to_e_k_dagger} and  \eqref{eq:bar_e_k+1_xi_to_e_k_xi}, we obtain \eqref{eq:error_dyn}.

Moreover, it turns out that the first $n \times n$ submatrix of $\breve{E}^\xi_{k+1}$ is the zero matrix, i.e.,
\begin{equation*}
\begin{split}
    \breve{E}^\xi_{k+1} &= {\mathbb I}_N - \begin{bmatrix}
    \mathsf{u}^\top \otimes I_n \\ {\mathbb W}^\top
    \end{bmatrix} {\mathbb 1}_N  \mathcal{H}^{-1}_{k+1} {\mathbb 1}_N^\top (K_{k+1}^{\mathsf{cons}})^{-1} \begin{bmatrix}
    \mathsf{u} & {\mathbb W}
    \end{bmatrix}\\
    &= \begin{bmatrix}
    0_{n \times n} & \check{E}^\xi_{k+1}\\
    0 & {\mathbb I}_{N-1}
    \end{bmatrix}
\end{split}
\end{equation*}
where $\check{E}^\xi_{k+1} = -\frac{N}{\sqrt{N}} \mathcal{H}_{k+1}^{-1} {\mathbb 1}_N^\top  (K_{k+1}^{\mathsf{cons}})^{-1} {\mathbb W}$.
Recalling from Lemma \ref{lem:est_rule} that $\breve{\Xi}_{k} =\begin{bmatrix}
I_n & \check{\Xi}_k\\
0 & \tilde{\Xi}_k \end{bmatrix}$, we obtain
\begin{equation}\label{eq:E_xi_Xi}
    \breve{E}^\xi_{k+1} (\breve{\Xi}_{k+1})^{l^*} =  \begin{bmatrix}
    0_{n \times n} & \check{E}^\xi_{k+1} (\tilde{\Xi}_{k+1})^{l^*}\\
    0 & (\tilde{\Xi}_{k+1})^{l^*}
    \end{bmatrix}.
\end{equation}
Thus, if $\alpha_{\lambda,i}$ is chosen such that $0<\alpha_{\lambda,i}<2/\bar \sigma^2$, then the matrix $E^\xi_{k+1} (\breve{\Xi}_{k+1})^{l^*}$ is Schur stable for $k \ge  k^*$ by Lemma \ref{lem:est_rule}, which completes the proof.
\bibliographystyle{abbrv}        
\bibliography{mybib_Aut}         

\end{document}